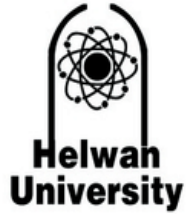

**Helwan University**

Faculty of Engineering
Biomedical Engineering Department

# Ultrasonic Estimation of Soft Tissue Visco Elastic Properties

A thesis submitted to the Faculty of Engineering, Helwan Univeristy in partial fulfillment
of the requirements for the degree of Master of Science

In

Biomedical Engineering

Submitted By

# Eng. Hassan Mostafa Ahmed Hassan Fahmy

B.Sc. Biomedical Engineering (2012)

## <u>Under supervision of</u>

**Prof. Mohamed Ibrahim El Adawy**
Department of Electronics, Communication and Computer Engineering,
Faculty of Engineering,
Helwan University

**Prof. Ahmed Farag Seddik**
Biomedical Engineering Department, Faculty of Engineering
Helwan University

**Assoc. Prof. Nancy Mustafa Ahmed Salem**
Biomedical Engineering Department, Faculty of Engineering
Helwan University

# Acknowledgement

I would first like to thank my thesis advisor Assoc. Prof. Nancy Mustafa Ahmed, Prof. Ahmed Seddik and Prof. Mohammed Ibrahim El Adawy, Faculty of Engineering Helwan University. Their office doors were always open whenever I fall into a trouble spot or had a question about my research or writing. They consistently allowed this thesis to be my own work, but steered me in the right direction whenever they thought I needed it.

I would also like to thank the experts who were involved in the validation survey for this research project: Eng. Prof. Mohammed Ali Maher. Without his passionate participation and input, the validation survey could not have been successfully conducted.

I would also like to acknowledge Prof. Ahmed Mohammed Ragab El-Bialy, Faculty of Engineering Cairo University and Prof. Mohammed Tarek Ibrahim El-Wakad Faculty of Engineering Helwan University as the second reader of this thesis, and I am gratefully indebted to them for their very valuable comments on this thesis.

Finally, I must express my very profound gratitude to my parents and to my colleagues specially Eng. Reda Abd El-Baset for providing me with unfailing support and continuous encouragement throughout my years of study and through the process of researching and writing this thesis. This accomplishment would not have been possible without them. Thank you.

Author
Hassan Mostafa Ahmed Hassan Fahmy Gabr
Biomedical department, Faculty of Engineering, Helwan University

# Publications

Hassan M. Ahmed, Nancy M. Salem, Ahmed F. Seddik, Mohamed I. El Adawy, "On Shear Wave Speed Estimation for Agar-Gelatine Phantom", International Journal of Advanced Computer Science and Applications (IJACSA), vol. 7, issue 2, pp. 401-409, 2016.

# Abstract

Conventional imaging of diagnostic ultrasound is widely used. Although it makes the differences in the soft tissues echogenicities' apparent and clear, it fails in describing and estimating the soft tissue mechanical properties. It cannot portray their mechanical properties, such as the elasticity and stiffness. Estimating the mechanical properties increases chances of the identification of lesions or any pathological changes. Physicians are now characterizing the tissue's mechanical properties as diagnostic metrics. Estimating the tissue's mechanical properties is achieved by applying an Acoustic Radiation Force Impulse (ARFI) on the tissue and calculating the resulted shear wave speed. Due to the difficulty of calculating the shear wave speed precisely inside the tissue, it is estimated by analyzing ultrasound images of the tissue at a very high frame rate.

In this study, the shear wave speed is estimated using finite element analysis. A two-dimensional model is constructed to simulate the tissue's mechanical properties. For a generalized soft tissue model, Agar-gelatine model is used because it has similar properties to the soft tissue. A point force is applied at the center of the proposed model. As a result of this force, a deformation is caused. Peak displacements are tracked along the lateral dimension of the model to estimate the shear wave speed of the propagating wave using the Time-To-Peak displacement (TTP) method.

Also, the behavior of other phantoms (with different elastic moduli) has been investigated in this research. Five phantoms with different elastic moduli; 5.2KPa, 9.8KPa, 23.9KPa, 44.2KPa and 67.3KPa respectively; have been investigated and their corresponding shear wave speeds have been estimated using the lateral Time-To-Peak displacement method.

The behavior of these phantoms to the shift of the excitation frequency of the pushing pulse has been investigated as well. This leads to the conclusion that the relationship between the peak displacement and the excitation frequency is inversely proportional.

A case study of liver and breast is investigated by constructing models for both the liver and breast tissues and applying an ARFI and estimating the resulting shear wave speed and comparing it with clinical results.

Experimental results are obtained so that the shear wave speeds are 1.13, 1.46, 2.28, 3.16 and 3.86 (m/sec.) compared to theoretical values of 1.32, 1.81, 2.82, 3.84 and 4.74 (m/sec.) for 5.2, 9.8, 23.9, 44.2 and 67.3 KPa phantoms respectively. Also results of case studies for breast tissue are obtained to be 2.0018-2.2853, 2.5191-3.9807 and 4.8363-7.4405 (m/sec.) compared to theoretical values of 2.3791-2.7472, 2.9673-4.5557 and 5.4944-8.7595 (m/sec.) for normal fatty, normal glandular and fibrous glandular breast tissues respectively. Liver tissue yields results of 0.4735-1.5179 and 1.8380-5.0223 (m/sec.) compared to theoretical values of 0.3546-1.3736 and 2.1718-5.6077 (m/sec.) for normal liver and cirrhosis liver tissue.

# Contents

# List of Tables

# List of Figures

# Abbreviations

| | |
|---|---|
| ARF | Acoustic Radiation Force |
| ARFI | Acoustic Radiation Force Impulse |
| FEA | Finite Element Analysis |
| FEM | Finite Element Modeling |
| FNM | Fast Near Method |
| MRI | Magnetic Resonance Imaging |
| ROE | Region Of Excitation |
| SWEI | Shear Wave Elasticity Imaging |
| SWS | Shear Wave Speed |
| SSI | Super Sonic Imaging |
| TTP | Time-To-Peak |
| VA | Vibro-Acoustograhy |

**Notation**

| | |
|---|---|
| [] | Rectangular matrix |
| {} | Column vector |
| $\delta_{ij}$ | Kroenecker delta |
| $\epsilon_{ijk}$ | Levi-Civita tensor |
| $\nabla$ | Differential operator |
| $\nabla^2$ | Laplacian operator |

**Greek symbols**

| | |
|---|---|
| $\alpha$ | Attenuation coefficient |
| $\theta_i$ | Incident wave angle |
| $\theta_r$ | Refracted wave angle |
| $\theta_t$ | Transmitted wave angle |
| $\mu$ | Lamè coefficient |
| $\boldsymbol{\lambda}$ | Lamè coefficient or wave length |
| $\eta$ | Viscosity constant |
| $\sigma_{ij}$ | Stress tensor |
| $\sigma'_{ij}$ | Deviatoric stress tensor |
| $\sigma_H$ | Hydrostatic stress |
| $\epsilon_{ij}$ | Strain tensor |
| $\epsilon'_{ij}$ | Deviatoric strain tensor |
| $\varepsilon$ | Uni-axial strain |
| $\sigma_t$ | Cauchy stress |
| $\varepsilon_l$ | Logarithmic |
| $\rho$ | Density |

| $\upsilon_o$ | Poisson's ratio |
| $\nu$ | Frequency |

# Chapter 1 | Introduction

## 1.1.    Introduction

Replacing healthy soft tissues by fibrotic tissues are the pathological change that may cause a malignant or benign tumor. The stiffness of these pathologic tissues is higher than the surrounding [1]. Elastic modulus is a measurand for the stiffness. It is the measure of the material's resistance to deformation in either compression or tension namely the elasticity modulus (E) and in shear namely the shear modulus (μ) [2]. Muscles and fibrous tissue are more resistant to deformation than other compliant tissues such as fat, due to their higher elastic moduli [3-5].

Deformation is a result of stress over the tissue. Formerly, it was manual palpation over the tissue. Nowadays, it is the beat or the push that is caused by the acoustic radiation force generated by the ultrasound probe; a procedure called Elastography procedure. This procedure can be accomplished by many methods. These methods are classified regarding either the obtained images or the source of excitation.

Regarding the obtained images, these methods are either qualitative; revealing relative stiffness differences; or quantitative; leading to an estimate for the underlying tissue elastic modulus using reconstruction methods. Also, classification regarding the source of excitation these methods are either external; which are the dynamic external methods and static external compression methods for mechanical excitation; or internal; like the physiological motion of the tissue itself or an Acoustic Radiation Force (ARF).

## 1.2.    Motivation and Objectives

Anatomical structures of soft tissues are obtained by traditional ultrasound imaging easily. Yet, imaging its mechanical properties is not of this feasibility. These mechanical properties are of significant importance as they give valuable information about the pathology of the soft tissue. Pathologic tissues exhibit stiffer behavior than other non-pathologic tissues. Thus, soft tissue stiffness is now considered as a metric for the tissue pathologic status. It is used as a diagnostic metric nowadays. Identification of the lesions or any other pathological changes in the soft tissue is more likely achieved by estimating its mechanical properties.

This has led to the interest about estimating the mechanical properties of soft tissue as it increases chances of identification of any lesion, mass or pathological change at early its stages.
Estimation of these mechanical properties; such as stiffness; is achieved by estimating the speed of shear wave that is induced inside the tissue itself. Shear wave speed is related to the stiffness of the soft tissue in which it is induced.

The objective of this research is to estimate the shear wave speed for agar-gelatin phantom and to investigate the displacement magnitude profile of the central node. Also, to estimate the shear wave speed for some case studies of breast and liver tissues.

## 1.3. Thesis Overview

In this research, a two dimensional Finite Element Model (FEM) is constructed simulating the soft tissue's mechanical properties. For a generalized model, Agar-gelatin model is used due to its similarity in behavior to soft tissue, where the agar is considered to act as scatterers and gelatin is considered to introduce the elasticity for the model.

A transient ARF is applied to the model to induce scatterers displacements. Then, peak displacements are tracked using Time-To-Peak displacement (TTP) method along the lateral direction starting from the focal zone and moving away towards either of the lateral edges. Speed; of the shear wave generated by the propagation of the peak displacements; is then estimated by calculating the time taken by the wave to reach a specific point on the lateral direction divided by the distance.

The two dimensional FEM has been constructed by using Finite Element Analysis (FEA) software LISA v8.0. The ARF has been simulated using the ultrasound pressure field simulation tool FOCUS. Calculated results are compared with the estimated results obtained from the shear wave speed estimation to calculate the deviance from the calculated expected results.

Also, five phantoms with different elastic moduli have been constructed and their response to the point ARFI is investigated. Their resulting Shear Wave Speeds are estimated using the lateral TTP method and are compared to the calculated results.

Case studies for breast and liver tissues have been introduced and their corresponding shear wave speeds have been estimated. Figure 1.1 shows a block diagram for the work concluded in this research.

There are four steps concluded by this research, the first step is to estimate the shear wave speed for agar-gelatin phantom so as to obtain the optimum frame rate that is proper for the software tools used throughout the thesis. The second step is to investigate the relationship of the central node's displacement magnitude profile with the shear wave speed estimated for five different elastic moduli phantoms in five preliminary experiments. The third step comprises using the conclusions obtained from the two previous steps to estimate the shear wave speed for two case study tissues; namely the breast and liver tissues. Finally the trade-offs introduced by shift of the stiffness value for a certain tissue are investigated.

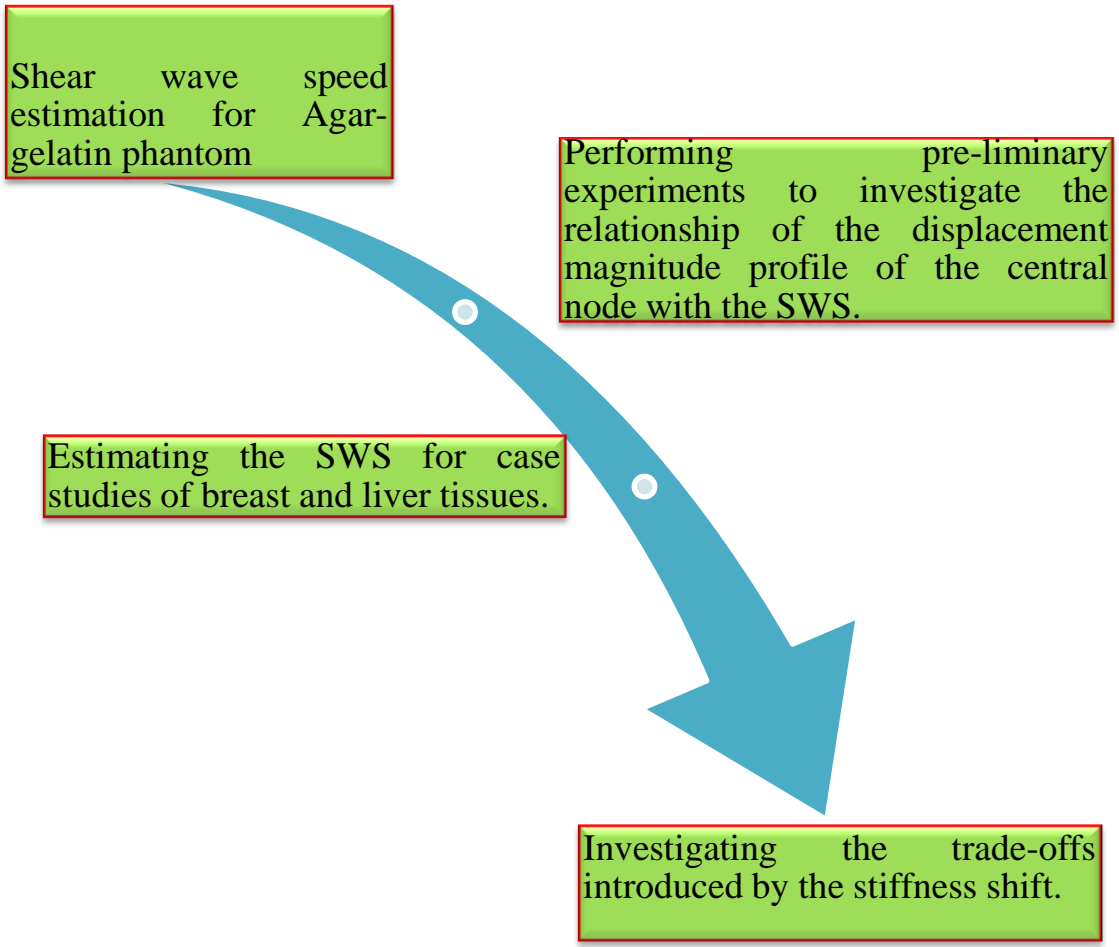

Figure 1.1: Block diagram for the work concluded in this research.

## 1.4.    Thesis Organization

This thesis consists of seven chapters and organized as follows:

**Chapter 1:** introduces the work done and describes the objectives, motivation and organization of the thesis.

**Chapter 2:** covers the relevant background about the ultrasound modality, the finite element modeling of a soft tissue as linear elastic and viscoelastic material and limitations of this modeling and the elastography imaging procedure.

**Chapter 3:** introduces the literature survey of the elastography imaging techniques and its application in the medical field.

**Chapter 4:** introduces the proposed method to measure the shear wave speed. The construction of the model and its calibration takes place. The simulation of the pressure field of the ARFI is obtained and the pushing sequences are generated. The application of this ARFI on the Agar-Gelatin FEM model and the behavior of the five different phantom models are covered.

**Chapter 5:** simulates the preliminary experiments and case studies. Preliminary experiments are carried out to study the displacement magnitude profiles. Two case studies have been simulated to investigate their behavior with respect to the application of the ARFI. These two cases are breast cancer and liver cirrhosis respectively.

**Chapter 6:** introduces the parameters affected by ARFI, such as Time-To-Peak displacement and Peak displacement magnitude.

**Chapter 7:** concludes the work in this research and remarking important points obtained from preceding chapters. Also introduces the future work that can be continued by the aid of this work. The idea of using the SWEI to detect the glaucomatous eye from non- glaucomatous eye and estimating the stiffness of the eye is introduced in this chapter. The feasibility of using such an imaging technique has been investigated; the limitations have been mentioned as well.

**Appendix A:** gives the detailed proof for the Hook's law for tensor form of the elastic material.

**Appendix B:** gives an overview about the deviatoric and dilatational parts for the stress tensor of the elastic material.

**Appendix C:** gives an overview about the parameters related to the ARFI such as the time to peak displacement and peak displacement versus the stiffness.

# Chapter 2 | Background

## 2.1.    Introduction

This chapter gives the necessary background to follow the work done in this research. Firstly, the ultrasound imaging fundamentals are covered. The second section will cover the Finite Element Modeling (FEM) theory. Then, the elastography imaging fundamentals and its methods are covered. Finally, Field II and Fast Near Method (FNM) are reviewed briefly.

## 2.2.    Ultrasound Imaging Fundamentals

Sound is defined as the mechanical energy transmitted by pressure waves in a material medium [6]. Sound moves from one location to another as it carries energy; that is; it can cause slight back and forth displacements of objects in its path. Source of sound wave in the medical ultrasound is a piezoelectric transducer, where the applied electrical signal on the transducer is electrical impulse.

Sound waves travelling inside the soft tissues are longitudinal waves, where the direction of propagation is the same direction of particles' vibration. It is the longitudinal waves which are important in ultrasound imaging due to the feasibility in propagation in soft tissues.

Longitudinal waves are characterized by their compressions and rarefactions, where at places of compressions the particles of the medium are pushed together and the density of the material at these places is slightly higher than it would be in the absence of the wave. At the places of rarefactions the particles of the medium are apart from each other leading to a slightly lower density of the medium material at these locations.

Strength of a sound wave is frequently quantified by quantifying the acoustic pressure. In a sound wave, at places where compressions take place there is an elevation in the pressure compared to the atmospheric pressure, and vice versa with rarefactions. The acoustical pressure is measured in Pascal (Pa).

### 2.2.1. Frequency of sound wave

The sound wave like any other wave is characterized by the frequency and the period of the propagating wave. Frequency is defined as the number of complete cycles per second [6]. In the case of sound wave, it is defined as the number of times per second the disturbance is repeated at any fixed and specified point along the path of the propagating wave and is measured by Hertz (Hz). Simply it is the number of complete oscillations made by the sound source. Period is defined as the time taken by one complete oscillation or time taken by one complete disturbance to repeat itself and is measured by Second (Sec.). Frequency and period (periodic time) are related to each other by:

$$\nu = \frac{1}{T} \qquad (2.1)$$

Where $\nu$ is the frequency of the propagating sound wave, and $T$ is the periodic time of the same wave. It is clear that the frequency is inversely proportional to the periodic time.

A classification scheme for the acoustic wave according to their frequency is shown in Fig. 2.1 [7]. Mostly audible sound wave can carry frequencies ranging from 20 Hz to 20 KHz, also termed as infrasonic range. Ultrasound ranges from 20 KHz to 1 MHz, also termed as ultrasonic range. Frequencies over than 1 MHz forms the waves used in diagnostic medical ultrasound.

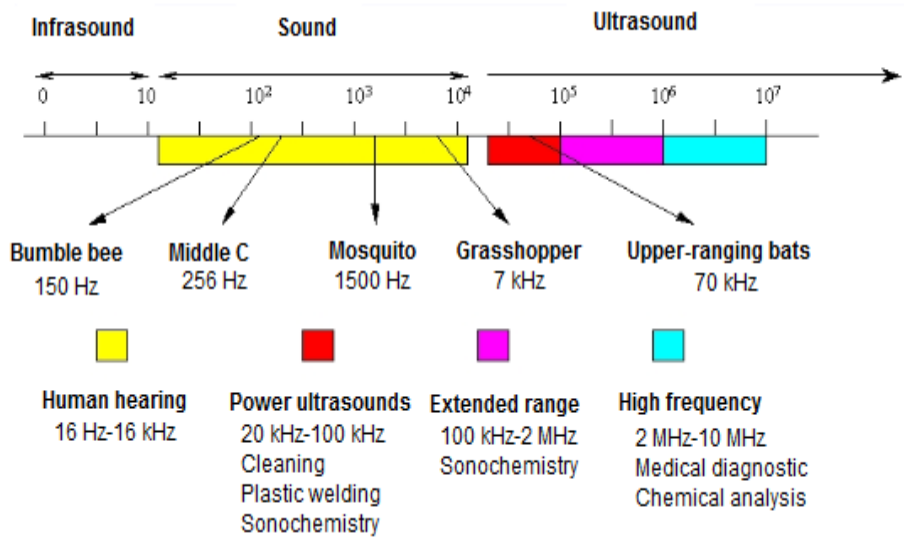

Figure 2.1: Classification scheme for the acoustic waves according to their frequency [7].

## 2.2.2. Speed of sound in medium

Speed of sound inside the medium is a characteristic for that medium. It depends on the medium characteristics like the Bulk's modulus of that medium. Specifically; for the longitudinal sound waves in either liquids or viscoelastic body tissues, the speed of sound is calculated by:

$$C = \sqrt{\frac{B}{\rho}} \qquad (2.2)$$

Where **C** is the speed of sound inside the tissue and is measured in $(\frac{m}{Sec.})$, **B** is the Bulk's modulus of that tissue and it is a characteristic for the tissue, and **ρ** is the density of the tissue measured in $(\frac{Kg}{m^3})$. From that equation, the speed of sound depends on the stiffness of the medium and on its density.

### 2.2.3. Wavelength of sound wave

Wavelength of the acoustic wave is depending on both the frequency $\boldsymbol{\nu}$ and the speed $\mathbf{C}$ of the sound wave propagating inside the soft tissue. It is defined as the distance between two successive compressions or two successive rarefactions [6]. It is given by:

$$\lambda = \frac{C}{\nu} \qquad (2.3)$$

Where $\boldsymbol{\lambda}$ is the wavelength, $\boldsymbol{\nu}$ is the frequency and $\mathbf{C}$ is the speed of the incident wave respectively. It is the speed of sound inside the soft tissue divided by the frequency of the incident sound wave on that soft tissue.

Higher frequencies results in smaller wavelengths and vice versa. The wavelength of the incident wave on a soft tissue where the speed equals $1540 \ (\frac{m}{Sec.})$ is given by:

$$\lambda = \frac{1.54 \ (mm)}{\nu \ (MHz)} \qquad (2.4)$$

Wavelength is the acoustical yardstick where it has a direct relationship with the spatial resolution of the ultrasound beam, as will be mentioned later. Briefly clarifying this relationship, the size of the object is significant when compared to the wavelength of the incident beam. Objects are either larger or smaller than the incident wavelength.

### 2.2.4. Intensity of sound wave

Acoustic pressure amplitude and acoustic intensity are also related to each other. Acoustic amplitude is defined as the maximum increase or decrease relative to the ambient conditions in the absence of the sound wave [6]. Acoustic intensity $\mathbf{I}$ is proportional to the square of the acoustic pressure amplitude $\mathbf{P}$, and the relationship is given by:

$$I = \frac{P^2}{2\rho C} \qquad (2.5)$$

Where **I** is the intensity of the incident beam, **P** is the pressure amplitude, **ρ** is the density of the medium and **C** is the sound speed in that medium.

Another issue should be highlighted which is the acoustic impedance **Z** for a medium. Acoustic impedance is a property for the medium or the tissue being imaged. It influences the amplitude or the strength of the reflected echo. It is the product of the medium's density **ρ** and its speed of sound **C** [6]. It is given by:

$$Z = \rho C \tag{2.6}$$

Acoustical impedance is measured by (Kg/m$^2$/sec.) or in rayls, where 1 rayl equals to 1 Kg/m$^2$/sec.

Sound speed in different biological tissues is measured and tabulated in Table 1 [1-2]. It is clear that the average speed of sound in soft tissue is equivalent to 1540 m/sec [6].

Table 2.1: Sound speed in different biological tissues [6].

| Material | Density ρ (Kgm$^{-3}$) | Speed C (msec.$^{-1}$) | Characteristic impedance Z (Kgm$^{-2}$sec.$^{-1}$)x10$^6$ | Absorption coefficient α (dB.cm$^{-1}$) @ 1MHz |
|---|---|---|---|---|
| Water | 1000 | 1480 | 1.5 | 0.0022 |
| Blood | 1060 | 1570 | 1.62 | -0.15 |
| Bone | 1380-1810 | 4080 | 3.75-7.38 | 14.2-25.2 |
| Brain | 1030 | 1558 | 1.55-1.66 | -0.75 |
| Fat | 920 | 1450 | 1.35 | -0.63 |
| Kidney | 1040 | 1560 | 1.62 | - |
| Liver | 1060 | 1570 | 1.62 | - |
| Lung | 400 | 650 | 0.26 | -40 |
| Muscle | 1070 | 1584 | 1.65-1.74 | 0.96-1.4 |
| Spleen | 1060 | 1566 | 1.65-1.67 | - |

At the incidence of the sound beam on an interface made by two materials having different acoustic impedances, an energy loss will be occured. Some of the incident energy will be reflected and the remainder will be transmitted. Reflected energy depends on the difference in the acoustic impedance between these two materials forming the interface. Hence, there are two cases of incidence, the perpendicular and the oblique incidence.

Firstly, the normal incidence will be discussed, where a specular surface; of dimensions much larger than the wavelength of the incident wave; is put along the path of the incident sound wave. The ratio between the reflected pressure amplitude $P_r$ to the incident pressure amplitude $P_i$ is called amplitude reflection coefficient **R,** and it is given by:

$$R = \frac{P_r}{P_i} = \frac{Z_2 - Z_1}{Z_2 + Z_1} \qquad (2.7)$$

Where $Z_1$ is the impedance of the proximal side and $Z_2$ is the impedance of the distal side.

The reflected wave amplitude depends on the difference in impedances between the two materials forming the interface.
Secondly; the oblique incidence; where it includes the propagation of the incident wave through the interface with some angle $\theta_t$ and reflection at some angle $\theta_r$ equal to the incidence angle $\theta_i$ in the opposite direction. Figure 2.2 illustrates clearly the reflection of an incident sound beam in both cases [3].

Another point that should be highlighted; the 3-dB rule; which states that a 3-dB increase (decrease) in the intensity and the power multiples (divides) the intensity and power by two.

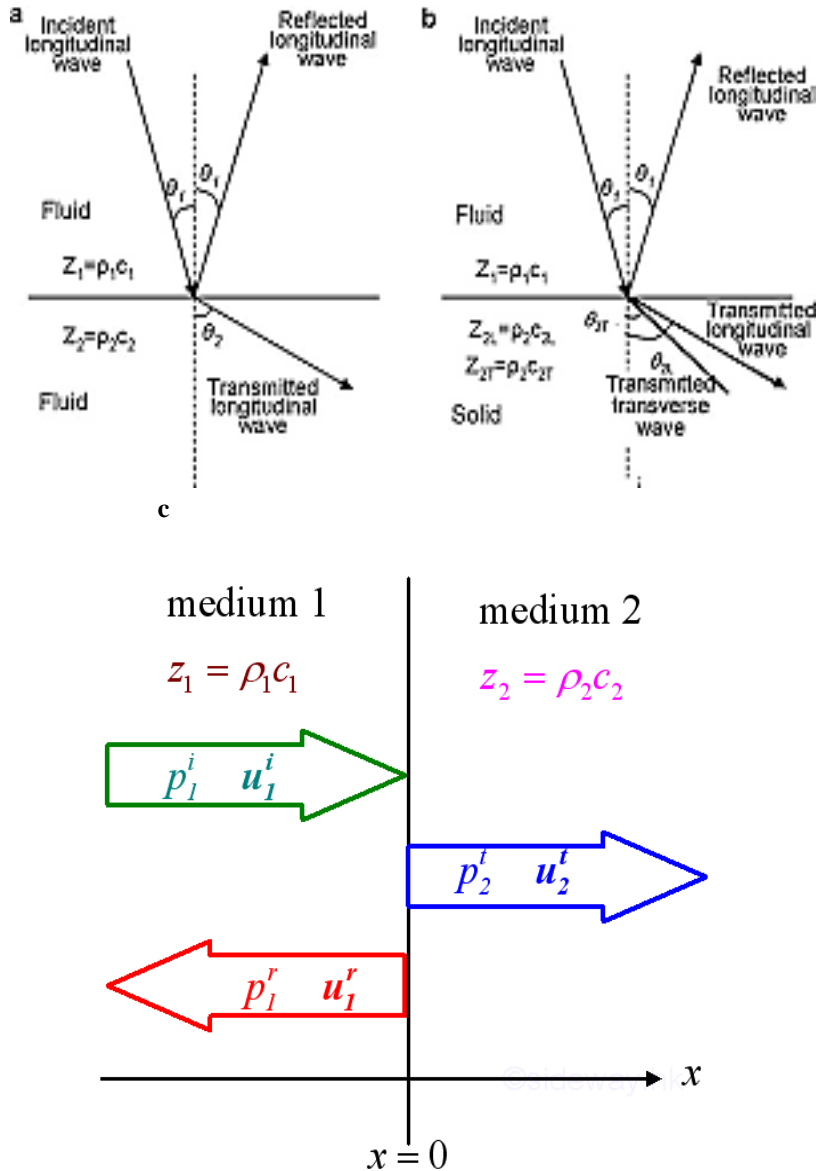

Figure 2.2: Sound beam reflections at medium boundaries: (a) Fluid-fluid interface, (b) Fluid-solid interface, and (c) Normal incidence.

### 2.2.5. Attenuation of sound wave

Attenuation of ultrasound means the degradation of the amplitude of the propagating wave. There are two sources for ultrasound attenuation inside the body, reflection and scattering at interfaces, and absorption. Attenuation coefficient **α** measures the degree by which the sound beam is attenuated in a tissue, and is measured by (dB/cm).

Attenuation of ultrasound beam depends also on the frequency of the incident beam, and this dependency is significant and in most cases is proportional to the frequency as shown in Fig. 2.3 [8]. From the relationship for liver tissue, it is clear that the increase in frequency of the incident beam results in the increase of the attenuation and hence the penetration is poor for deep tissues. The attenuation is calculated as the attenuation coefficient multiplied by the distance traveled by the beam, as given by:

$$\text{Attenuation (dB)} = \alpha \text{ (dB/cm)} \times D \text{ (cm)} \qquad (2.8)$$

Where **D** is the depth at which the attenuation is calculated.

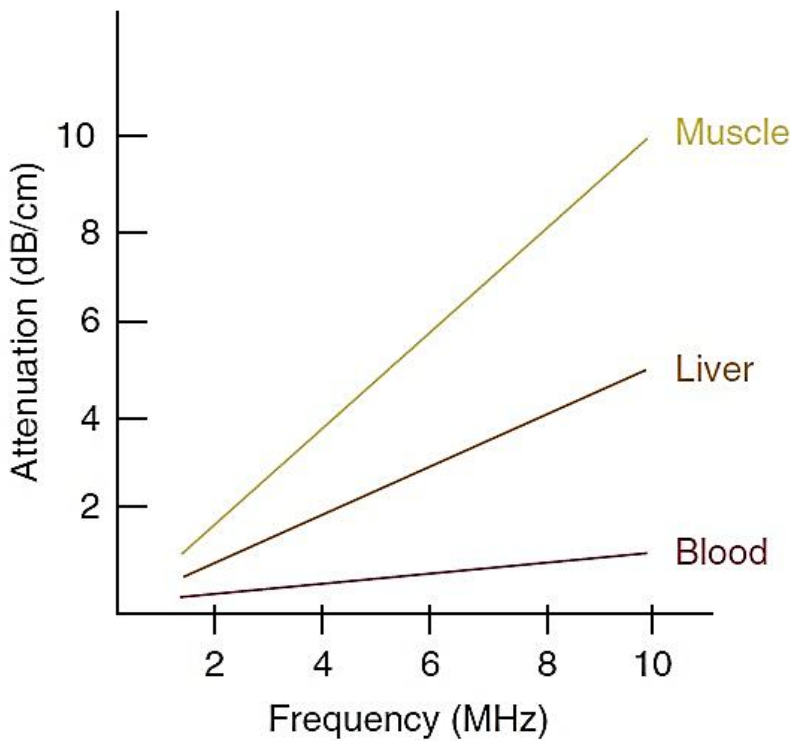

Figure 2.3: Attenuation of ultrasound beam versus frequency of the incident beam for some biological tissues[8].

## 2.2.6. Resolution of ultrasound beam

Spatial resolution refers to how much closely two reflectors are positioned to each other and still be differentiated by the ultrasound beam as two separated reflectors on an image. Spatial resolution is primarily divided into axial and lateral resolutions.

### Axial Resolution

Axial resolution is the minimum distance between two reflectors along the beam's direction after which these two reflectors will be distinguished as one complete reflector on the display as illustrated in Fig. 2.4. Pulse duration of the pulse transmitted to the medium determines the axial resolution, which is the number of complete cycles multiplied by the periodic time of one complete cycle. It is given by:

$$PD = N_c \times T \qquad\qquad (2.9)$$

Where **PD** is the pulse duration measured in seconds, $N_c$ is the number of complete cycles and **T** is the periodic time of the pulse transmitted to the medium and it is measured by seconds. Another issue arises here, which is the dependency of the axial resolution on the frequency of the incident wave, where there is an equivalent relationship between the **PD** and the frequency, and it is given by:

$$PD = \frac{N_c}{\nu} \qquad\qquad (2.10)$$

From the preceding equation, the axial resolution depends on the frequency, where higher frequencies lead to smaller **PD** which results in a better axial resolution, and vice versa. Also, higher frequencies lead to poorer penetration according to the attenuation-frequency curve. It is a trade-off between better resolution and deeper imaging.

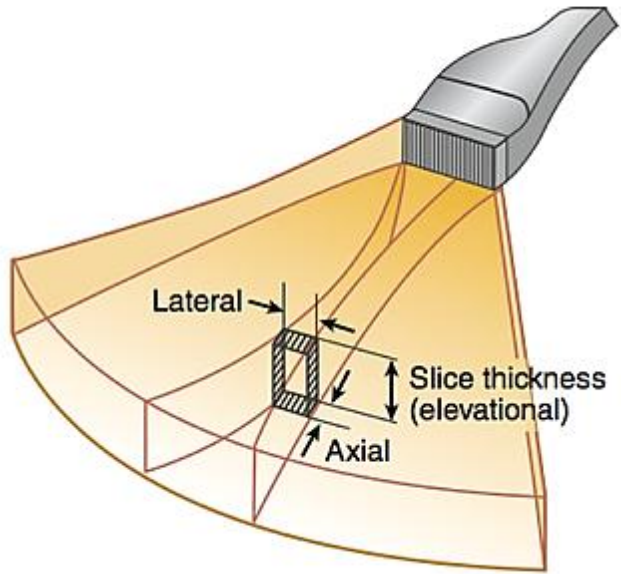

Figure 2.4: Ultrasound transducer resolution.

**Lateral Resolution**

Lateral resolution, is the minimum spacing between two reflectors laterally after which these two reflectors cannot be distinguished as two separate reflectors. Lateral resolution is determined by the beam width as shown in Fig. 2.5. Beam width at the focal zone is calculated by:

$$W = \frac{1.22\lambda}{d} \tag{2.11}$$

Where **d** is the diameter of the transducer, i.e. the diameter of the incident beam.

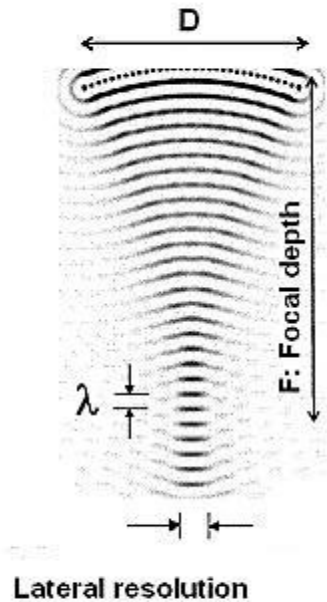

Figure 2.5: Lateral resolution of ultrasound beam.

## 2.2.7. Transducer array types

The ultrasound transducers are made up of arrays, whether linear, curvilinear or phased arrays. The array transducers have many advantages, amongst these advantages; enabling the electronic beam steering, electronic beam focusing and beam forming providing effective control of the focal distance and the beam width throughout the imaging field.

Array transducers assembly consists of a group of closely spaced piezoelectric elements, each with its own electrical connection to the ultrasound instrument. This design enables the elements to be individually excited or to be excited in groups. Also it enables the amplification of each echo signal individually before being combined together to form the echo beam. Figure 2.6 shows the types of the array transducer and Fig. 2.7 shows the electronic beam focusing.

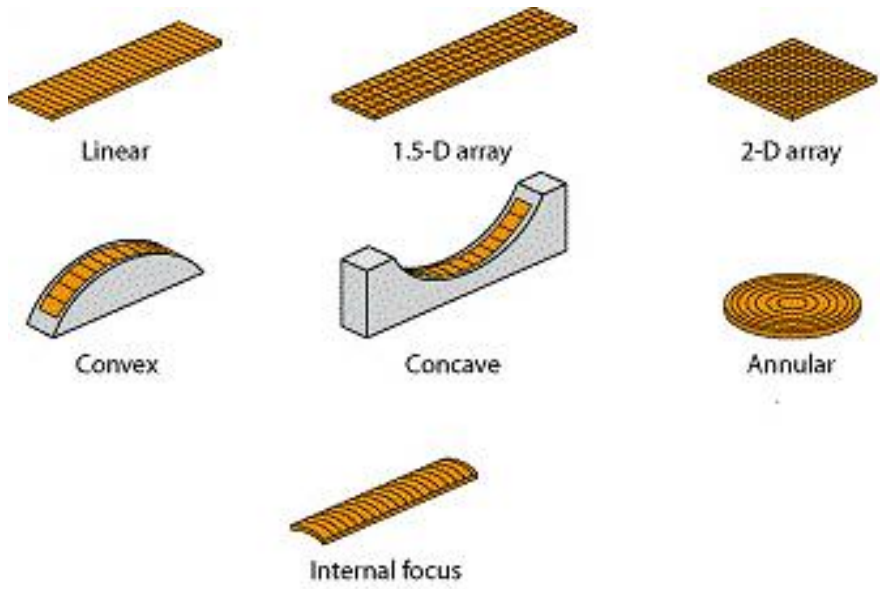

Figure 2.6: Array transducer types.

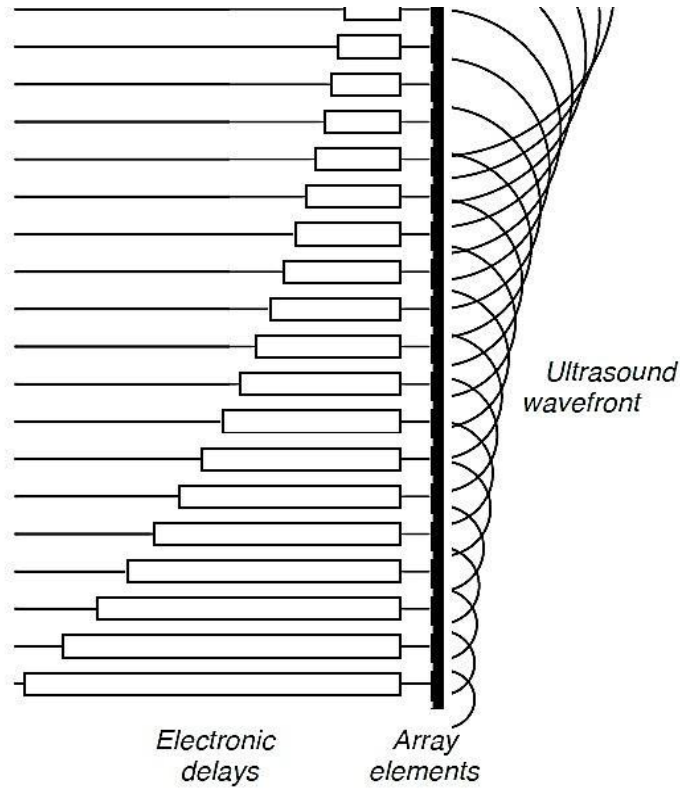

Figure 2.7: Electronic focusing of ultrasound beam.

## 2.3. Finite Element Modeling (FEM)

This section discusses the proof of the elastic material model and the improvement added by the viscosity term to the elastic model to become viscoelastic model.

If a material is assumed to be reversible and path independent, it is said to be an **elastic material**; this means that:

  a. a deformed body recovers to its initial configuration when forces causing deformation is removed.
  b. there is a unique relation between stresses and strains in the body.

Now we turn for the concept of **linear elastic materials**, and it is said so if and only if there is a linear relationship between the stress and the strain for that material.

Hence we can describe the linear elastic material by using the generalised Hook's law on tensor form [4], and it is given by:

$$\sigma_{ij} = C_{ijkl}\epsilon_{kl} \tag{2.12}$$

Where $\boldsymbol{\sigma_{ij}}$ is the $2^{nd}$ order stress tensor, $\boldsymbol{\epsilon_{kl}}$ is the $2^{nd}$ order strain tensor and $\boldsymbol{C_{ijkl}}$ is the $4^{th}$ order tensor governing the elastic material constants and having 81 components. For the detailed proof refer to appendix A.

Focusing on simplifying the $4^{th}$ order stiffness tensor as it governs the elastic material constants and having 81 components, simplifying these components to reduce the complexity of the calculation. Introducing stress and strain symmetry for Eqn. 2.12 to reduce the $4^{th}$ order stiffness tensor to a 36 independent components, where $\boldsymbol{\sigma_{ij}} = \boldsymbol{\sigma_{ji}}$ and $\boldsymbol{\epsilon_{ij}} = \boldsymbol{\epsilon_{ji}}$. This is called **minor symmetry**.

By differentiation of Eqn. A.12 with respect to $\boldsymbol{\epsilon_{kl}} = \boldsymbol{l_{kl}}$, the **major symmetry** is reached, and is given by:

$$C_{ijkl} = \frac{\partial^2 u}{\partial l_{ij}\,\partial l_{kl}} = C_{klji} \tag{2.13}$$

This is the major symmetry of the elastic tensor, leading to the advantage of reducing the number of components to be 21 independent components / coefficients. Further assuming of the material to be isotropic leads to describe the elastic coefficients using only two material constants, namely Lamè elastic constants, $\boldsymbol{\mu}$ and $\boldsymbol{\lambda}$, and now the $4^{th}$ order stiffness tensor can be given by:

$$C_{ijkl} = \lambda\,\delta_{ij}\delta_{kl} + \mu\,(\,\delta_{ik}\delta_{jl} + \delta_{jk}\delta_{il}\,) \tag{2.14}$$

By substituting in Eqn. A.12, we obtain the following:

$$\sigma_{ij} = \lambda\,\epsilon_{kk} + 2\,\mu\,\epsilon_{ij} \tag{2.15}$$

But the relation between these constants and the Young's modulus and Poisson ratio ($\mathbf{E}, \upsilon$) is given by Eqns. 16 and 17.

$$\mu = \frac{E}{2\,(1+v)} = G \qquad\qquad (2.16)$$

$$\lambda = \frac{vE}{(1+v)\,(1-2v)} \qquad\qquad (2.17)$$

Hence, the equation of the stress-strain is given by:

$$\sigma_{ij} = \frac{vE}{(1+v)(1-2v)}\,\epsilon_{kk}\delta_{ij} + \frac{E}{(1+2v)}\,\epsilon_{ij} \qquad\qquad (2.18)$$

Now the number of unknown elastic coefficients has been reduced to two coefficients, namely, the $\mathbf{E}$ and $\upsilon$. The detailed proof is introduced in appendix B.

In the following part further decomposition of both the stress and strain tensors, and the contribution of each constituent will be investigated. Each of the stress and the strain can be decomposed into **deviatoric** and **dilatational** parts. For the stress tensor, the equation will be:

$$\sigma_{ij} = \sigma'_{ij} + \frac{1}{3}\left(\sigma_{kk}\delta_{ij}\right) = \sigma'_{ij} + \sigma_H\delta_{ij} \qquad\qquad (2.19)$$

Similarly for the strain, the equation will be:

$$\epsilon_{ij} = \epsilon'_{ij} + \frac{1}{3}\left(\epsilon_{kk}\delta_{ij}\right) = \epsilon'_{ij} + \frac{1}{3}\left(\epsilon_v\delta_{ij}\right) \qquad\qquad (2.20)$$

Moreover, we can define the relationships between both the deviatoric stresses and strains, and the dilatational stresses and strains, to be given by Eqn. 2.21 and 2.22.

$$\sigma_{ij} = 2G\,\epsilon'_{ij} \qquad\qquad (2.21)$$

$$\sigma_H = K\,\epsilon_v \qquad\qquad (2.22)$$

Where **K** and **G** are given by:

$$\mu = \frac{E}{2\,(1+v)} = G \qquad\qquad (2.23)$$

$$K = \frac{E}{3(1-2v)} \qquad\qquad (2.24)$$

The isotropic part of the stress-strain tensor is represented by the hydrostatic stress $\sigma_H$ and the volumetric strain $\epsilon_v$, while the states of the stress-strain with zero dilatation are represented by the deviatoric stress $\sigma'_{ij}$ and deviatoric strain $\epsilon'_{ij}$ tensors, and the deviatoric part here means the volume is preserved. This decomposition is useful when we introduce the viscosity definition, because the time dependent response is assumed to be governed by the deviatoric stress.

The elastic properties of soft tissues can change depending on the rate of loading and deformation, which means that soft tissues are rate dependent. Also, for a given constant load or deformation, we may experience creep or stress relaxation which means that the elastic properties of soft tissues are time dependent even when the rate of loading or deformation is zero. This rate and time dependence is caused by internal friction, such as friction between cells in the soft tissue. In order to capture the rate- and time dependent response of the soft tissue, we need to include viscoelastic material properties. In general, the viscoelastic relation between stresses and strains can be nonlinear. However, we limit the discussion to linear viscoelasticity.

In the case of a linear viscoelastic material, the relation between state-of-stress and state-of-strain in the material is linear. Rheological models can be used to describe viscoelastic materials. The mechanical elements used in a viscoelastic rheological model are linear springs and viscous dashpots, which are shown in Fig. 2.8.a and 2.8.b.

The stress-strain relation for the two rheological elements can be written as:

$$\sigma = E\varepsilon \tag{2.25}$$
$$\sigma = \eta\dot{\epsilon} \tag{2.26}$$

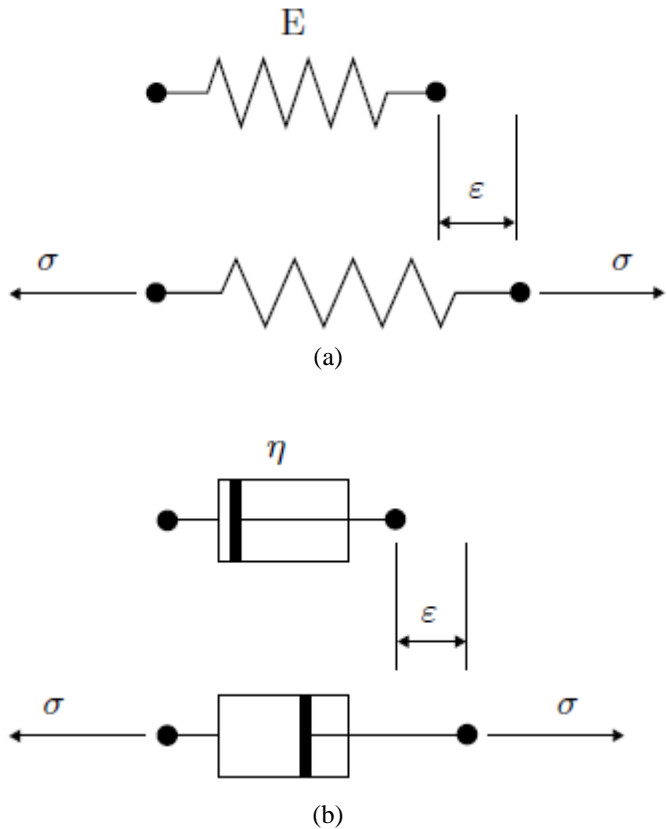

(a)

(b)

Figure 2.8: Rheological elements used to describe the viscoelastic material. (a) Linear spring element. (b) Linear viscous dashpot element.

A generalized Maxwell model is a combination of several Maxwell elements and a spring in parallel. The viscoelastic response of soft tissues is dependent upon the frequency of the loading. Hence, it is important to include Maxwell elements that can account for both the high-frequency response and the low-frequency response. A general higher-order Maxwell model is illustrated in Fig. 2.9, where N denotes the number of Maxwell elements used in the model. We note that a generalized Maxwell model will be included in the Finite Element Method (FEM) model in chapter 4. Thus, a brief discussion of the generalized Maxwell model is appropriate.

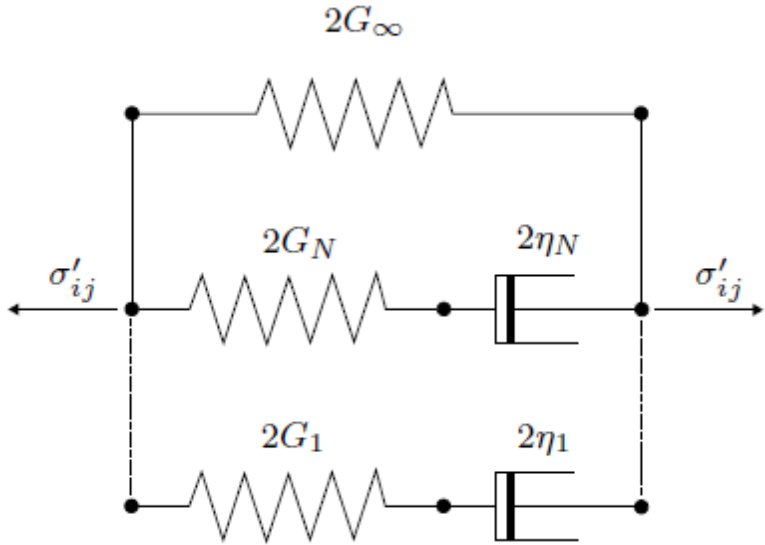

Figure 2.9: Rheological model of a generalized Maxwell material model, Where N is the total number
of Maxwell elements used in the model.

## 2.4. Elastography Imaging Fundamentals

Elasticity imaging methods adopt one concept which is applying either a mechanical excitation or stress to tissues. Stress can be external or internal excitation source. The internal source may be the physiological motion of the tissue itself or an acoustic radiation force (ARF). The resulting tissue deformation (displacement) is measured in response to that excitation using ultrasound, magnetic resonance or optical methods [1].

Based on the Helmholtz model for shear wave propagation; introduced by Eqns. 2.27, 2.28 and 2.29; the measured tissue deformation can be related to tissue stiffness.

$$\mu \nabla^2 u - \rho \left( \frac{\partial^2 u}{\partial t^2} \right) = \text{zero} \qquad (2.27)$$

$$C = \sqrt{\frac{\mu}{\rho}} \qquad (2.28)$$

$$\mu = \frac{E}{2(1+v)} \qquad (2.29)$$

Where; $\mu$ is the shear elasticity modulus.

When the imaging methods were first proposed, the excitation source obtained and considered as the physiological pulsation of the tissue itself, and ultrasound was used to monitor the tissue response [9, 10] where correlation between obtained A-scans is used to estimate small displacements caused by tissue pulsation [11, 12]. Then, dynamic methods were introduced, where dynamic external vibration is used to create shear waves inside the tissue to be studied (Sonoelasticity) [13] and methods using external static compression for mechanical excitation (strain imaging) [14]. The excitation using Acoustic Radiation Force Impulse (ARFI) was introduced in the early 90s by Sugimoto *et al.* [15]. This method has an advantage of coupling the source of excitation to the organ under study directly, rather than being coupled through intervening tissues. Greenleaf *et al.* and Parker *et al.* provide efficient reviews of elasticity imaging methods [16, 17]. Nyborg *et al.* [18] introduced a model of the tissue to be acting as a viscoelastic fluid in response to the ultrasonic wave propagation, and under plane wave assumptions, the ARFI is given by:

$$F = \frac{2\alpha I}{C} \qquad (2.30)$$

where $\boldsymbol{\alpha}$ (dB/cm.MHz) is the acoustic absorption coefficient of the tissue, $\mathbf{I}$ is the temporal average intensity of the wave and $\mathbf{C}$ (m/sec.) is the speed of sound in tissue.

Besides, the contribution of scattering is neglected because the majority of attenuation arises from absorption [18, 19]. The relationship between the depth and the frequency should be considered. The attenuation increases by increasing depth for higher frequencies. As a result, there is an optimal frequency for each depth which depends on the attenuation-frequency tradeoff.

The ARFI can be applied for different temporal duration's methods, such as the quasi-statically, transient method, and harmonic method [19]. The Quasi-static method proposes that the excitation pulses are applied on the tissue to reach a steady state response, typically longer than one second. The transient method proposes that the excitation is applied for a very short duration, typically a temporal impulse, faster than the natural frequency of the tissue associated with the dynamic tissue response. The harmonic method proposes the application of the impulse in a harmonic way; a pulsed manner; to achieve a sinusoidal tissue excitation of one or more frequencies. The procedure of elastography imaging is shown in Fig. 2.10.

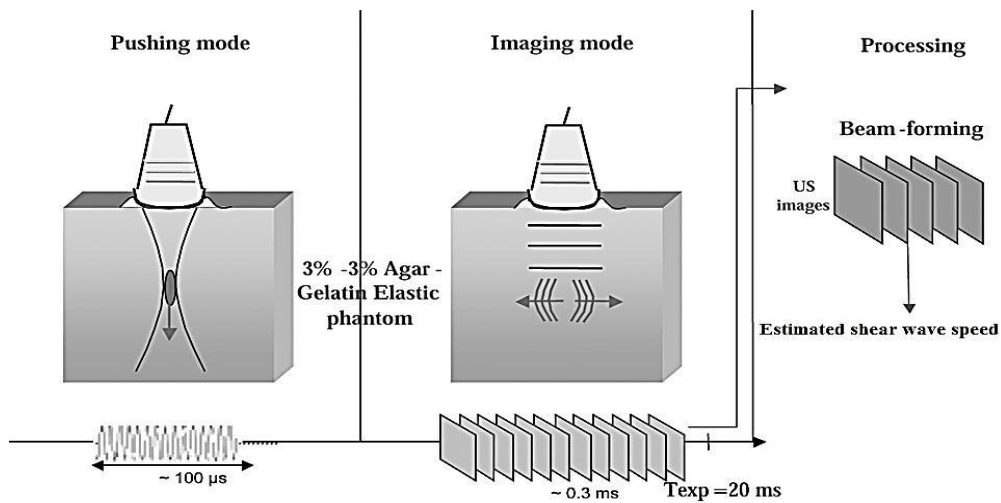

Figure 2.10: Elastography imaging procedure [20].

## 2.5. Field II

Field II is a program for simulating ultrasound transducer fields and ultrasound imaging using linear acoustics. Tupholme-Stepanishen method is used for calculating pulsed ultrasound fields. Calculating the emitted and pulse-echo fields for both the pulsed and continuous wave case for a large number of different transducers is achievable with this program. Also any kind of linear imaging can be simulated as well as realistic images of human tissue. The program is running under Matlab on a number of different operating systems [21]. Figure 2.10 [20] shows the output of Field II program simulating ultrasound probe image of a synthetic kidney.

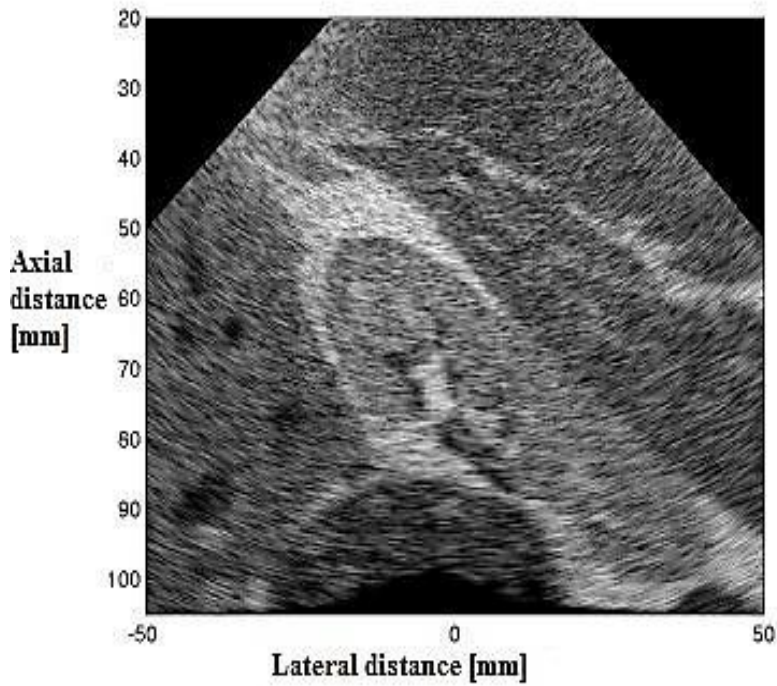

Figure 2.11: Synthetic kidney B-mode image.

## 2.6. Fast Near Method (FNM)

Fast and accurate evaluation of continuous wave pressure fields over large computational volumes is required for thermal therapy stimulations. The point source superposition is the standard approach. Each radiating aperture is approximated as an ensemble of point sources radiating in a homogeneous, linear medium. Yet conceptually simple, point source superposition is both inaccurate, especially in the near field region, and computationally expensive. The impulse response approach [22] was developed in the 1960's and 1970's for simple geometries, for example baffled circular and rectangular pistons, to alleviate these problems. The double integration required by point source superposition to a single integration is reduced by using the impulse response approach. Owing to the singular nature of the integrand the impulse response approach; like the point source superposition approach; is still inaccurate in the near field.

These problems are solved by removing the singularity from the impulse response expressions by the fast near field method (FNM) [22, 23]. For baffled pistons with simple geometries, a double integral is reduced to a rapidly converging single integral by FNM, then evaluated using Gauss quadrature. Gauss quadrature achieves exponential convergence within the nearfield region as the FNM integrand is smooth, thus generating machine accuracy with a small number of quadrature points. For a circular piston [24], the FNM utilizes a spatially band limited single-integral expression for all field points, while for the rectangular piston [25], the field is expressed as a sum of bandlimited single integrals. Expressions have also been developed for triangular pistons [26], apodized circular pistons [27], apodized rectangular pistons [27], and spherical shells [27]. Research is also being conducted on developing expressions for curved rectangular strips and continuum modeling of apodized arrays. The FNM, which converges much more rapidly than the point source superposition method or the impulse response, produces accurate numerical results in a fraction of the time required by other approaches. In addition, the FNM has been adapted to transient problems encountered in imaging applications. For time-domain problems, the FNM avoids the temporal aliasing problems associated with the impulse response approach while also providing a fast and efficient method for computing transient fields. For these reasons, FNM is chosen.

## 2.7.    Summary

In this chapter, the basics of the ultrasound imaging were reviewed. This includes definition for some ultrasound fundamentals like the frequency and the wavelength and speed of sound. Also attenuation and resolution have been defined. Finite element model proof is introduced and the detailed proof is referred to in appendix 1 along with the proof of the Lame constants. Elastography imaging fundamentals have been reviewed as well.

# Chapter 3 | Literature Survey

## 3.1. Introduction

Applying a specific stress (force) then measuring the corresponding mechanical response is considered as the most common method for elasticity imaging. By having variations in the duration of the applied force and different measuring methods, a variety of imaging methods have been obtained, each having advantages and drawbacks. The most common methods is reviewed in this chapter. These methods are shown in Fig. 3.1 [28], where the two major categories; namely the mechanical excitation methods and the ultrasound excitation methods; include the most common techniques used in elastography imaging. In mechanical excitation methods, the transient elastography is the most common technique, whereas in ultrasound radiation force excitation methods both the shear wave elasticity imaging (SWEI) and supersonic shear imaging (SSI) are the most common techniques.

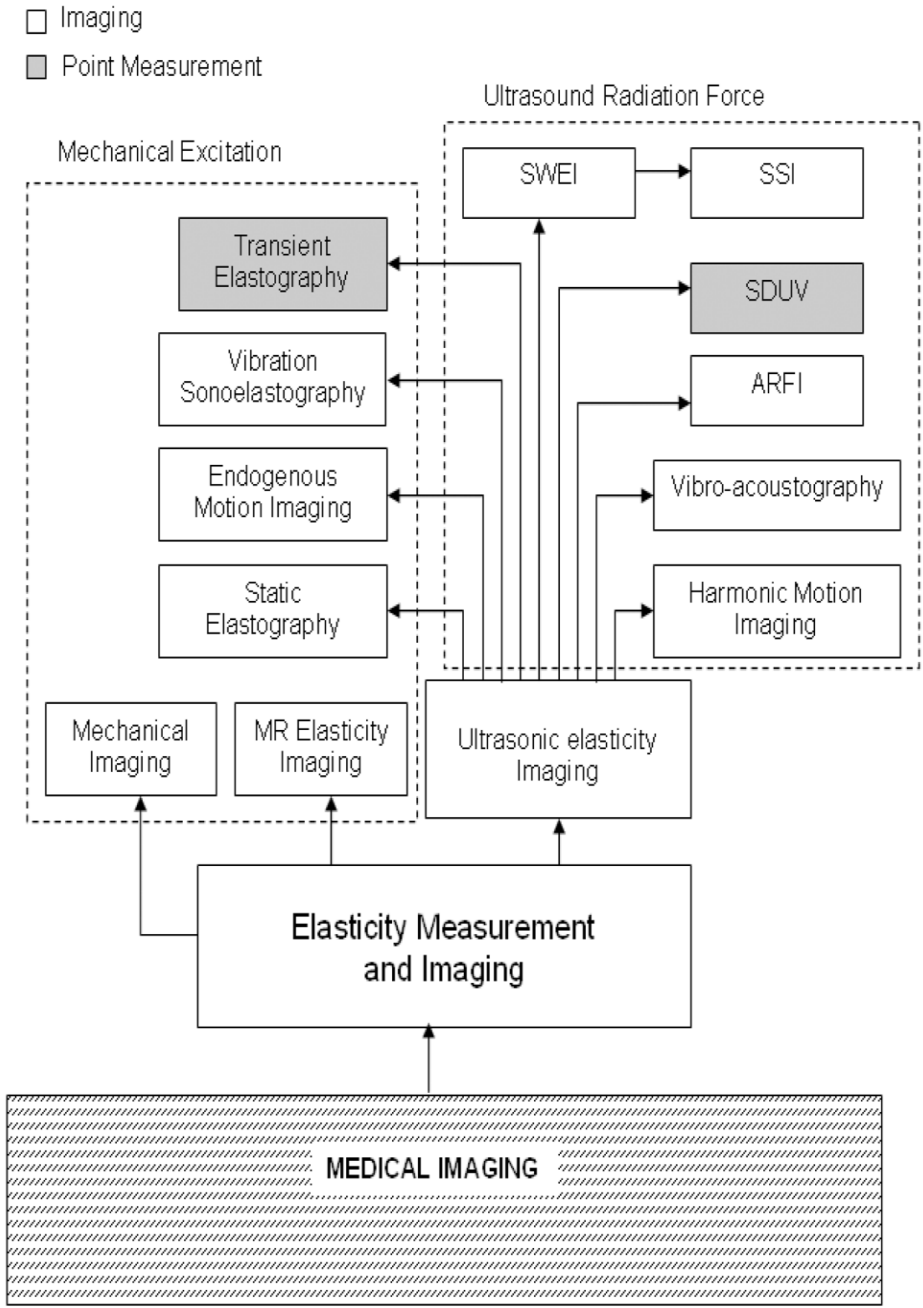

Figure 3.1: Block diagram of elasticity measurement and imaging and different methods included within this imaging modality. The techniques are categorized by their excitation method, mechanical or ultrasound radiation force [28].

## 3.2.  Survey

### 3.2.1.    Sonoelasticity

Harmonic shear waves are generated mechanically by using external actuators in contact with the skin. Doppler or any other imaging technique is used for measurement of wave propagation [29].

Krouskop *et al.* induced the shear wave inside the muscle tissue of thigh using a motorized actuator placed on the medial side of the thigh. The ultrasound transducer was put on the lateral side of the thigh to measure the wave propagation using Doppler methods [29]. On the other hand, Lerner *et al.* proposed the use of the acoustic horn to generate the waves in the phantom and the use of colored Doppler for measurement [30].

Shear wave velocities distribution is obtained by sonoelastography for phantoms, human prostates, and skeletal muscles [31].

### 3.2.2.    Acoustic Radiation Force Impulse

Acoustic radiation force impulse (ARFI) is defined as the force resulting from the momentum transfer from the propagating ultrasound wave through the tissue due to absorption and scattering mechanisms [16]. Displacements in tissue can be generated using a focused force impulse, these displacements (deformations) are relaxed and the tissue returns to its original position by the removal of the force [32].

In response to this focused force, the tissue within the region of excitation (ROE) is deformed and shear waves are generated and propagated away from the ROE. After a short duration excitation, an ordinary imaging procedure is performed to image these waves departure along the transverse direction of the ROE. Along a single line, excitation is accomplished and measurement is made, then, another adjacent line is investigated and so on till the image of tissue response is constructed. The tissue response is characterized by a set of parameters; the peak displacement, and the time the tissue takes to reach the peak displacement and the recovery time [33].

This method has been used for many applications. It is used with phantom tissue imaging like the agar-gelatin phantom [33], imaging thermally induced lesions [34], abdominal imaging of lesions [35], human prostate imaging [36] and imaging of cardiovascular vessels [37, 38].

### 3.2.3. Transient Elastography

An external source of vibrations is used to generate waves and to provide a single cycle of low frequency; typically 40-50 Hz. Compressional and shear waves are generated together by this sort of excitation. Fortunately, they are separated from each other by the time lag between them, as the compressional wave is much faster than the shear one [39, 40]. Transient excitations avoid biases caused by the sinusoidal excitations of cylindrical source [41, 42].

Motion tracking is the most important part of this method, where cross-correlation is used to locate the time shift between two echo signals, and by knowing the speed of sound in tissue, one can estimate how much motion has occurred inside the tissue.

Dutt *et al.* were pioneers to use mechanical actuation and to obtain a measurement of shear waves using this method [41]. Dutt *et al.* compared the estimated output of a magnetic resonance imaging for the shear wave with that estimated by ultrasound beam. Transient elastography can be used to measure stiffness in phantoms, breast, and skeletal muscle [42].

### 3.2.4. Shear Wave Elasticity Imaging

The modulated ultrasound beam that generates acoustic radiation force was not used until Nightengale *et al.* [43] have made their experiments on the proposed theory of Sarvazyan *et al.* [28] The generated force, when applied to the tissue, generates the shear waves that are detected by any other method. It is used to palpate the tissue but from the inside. Thereby, it has replaced the physicians' finger and is used as a virtual finger. Having high localization of induced strain (due to high attenuation after a few micrometers away from the ROE) was the main difference between this method and any other elasticity imaging method. The high absorption feature of these newly generated waves is the most important reason for the feasibility of using it. Then, an induced vibration is possible to be located in a very tiny portion of the tissue, namely in the focal zone.

In this technique, the major drawback is the small value of deformation. Hence, very complex signal processing methods are required to accurately estimate the motion [44-46]. This method has been used for investigation of phantoms, liver, prostate, and cardiac tissue [47, 48].

### 3.2.5.    Supersonic Shear Imaging

In the preceding methods, the imaging procedure was made along a single line and with single focal point probably the focal point. The supersonic shear imaging involves investigation of multiple focal points for the same line. The focal point is changed axially along the vertical beam with a speed much higher than the propagation speed of the shear wave resulted [49]. The multiple shear waves resulting from many focal locations (axially) constructively interfere to construct a conical shear wave [49].

Resulting in a Mach cone, where the Mach number of excitation can be adjusted to make the shear wave directionally oriented. It can be used for phantoms, liver, skeletal muscle and breast assessments [50].

It has been noticed that in viscous fluids, applying acoustic radiation force generates acoustic streaming or fluid flow [18], and this fluid flow has a velocity proportional to the fluid viscosity and the boundary condition. It is worthy to mention that phenomenon as it helped greatly in breast cancer detection and differentiation.

Starritt *et al*. were the first to investigate this phenomenon of generation of acoustic streaming [18]. Nightingale *et al.* [43] were the first to use it to differentiate between the fluid filled and solid lesions in the breast. This was achieved by interspersing pushing pulses with Doppler pulses to detect the resulting fluid flow using Doppler techniques [43].

### 3.2.6.    Quasi-static Elastography

Obtaining anatomy maps before and after inducing a small deformation of the tissue is considered the simplest approach to acquire elasticity information from soft tissue.

These anatomy maps are the Radio Frequency (RF) echo signals, where tiny motion (typically in the order of microns) induces a change in the phase of the RF echoes that can be tracked.

Amongst the common methods of tracking small changes of the RF signals are the correlation based methods, where they have the advantage of producing unbiased estimates of displacement with very low variance.

Relative strain images are built up from the displacement gradient values. This method of imaging is firstly introduced by Ophir *et al.* [51].
This method has contributed in many medical applications such as: breast imaging, prostate, thyroid, muscle and lymph nodes [52-55].

### 3.2.7.     Endogenous Motion Imaging

In this method, the excitation source for elasticity imaging can be the endogenous motion of the body itself. Inside the heart and the vascular system the excitation is considered as the pumping action of the heart. The pumping action of the heart muscle generates displacement waves in the cardiac tissue which characterizes the heart tissue material properties [56-60].
A high frame rate of ultrasound imaging is required to perform these measurements.

### 3.2.8.     Vibro-Acoustography(VA)

It is a method where the Acoustic Response of an object to the harmonic radiation force of ultrasound is used for imaging and material characterization [61, 62].

The procedure of imaging starts by focusing two ultrasound beams of slightly different frequencies at the same spatial location and tissue starts to vibrate as a result of the force exerted on it with a frequency equal to the difference of the primary frequencies of the incident beams.

The typical excitation frequency is in the range of 2-5 MHz resulting in a frequency difference of dozens Kilohertz (10-70 KHz), where the difference of the frequencies ($\Delta f$) is of magnitude of two orders less than the magnitude of the incident frequencies for ensuring that $\Delta f =<< f_1, f_2$. It is $\Delta f$ which is responsible for tissue vibration.

The Acoustic response is picked up by means of a hydrophone, and these co-focused beams are raster scanned across the object and the resulting acoustical signal is recorded.

Modulating the brightness of each pixel proportional to the acoustical signal amplitude from the excitation point of the object forms an object image.

Images resulting from this technique have some unique characteristics differentiating it from traditional ultrasound imaging. This is due the non-linear phenomenon of frequency conversion. A significant characteristic for this method over conventional imaging is obtaining speckle free images.

The ability to image specular surfaces regardless the orientation of the probe is another significant advantage, where in conventional imaging the specular surfaces are imaged when the probe is perpendicular to the surface.

This technique has contributed in the medical field in many applications such as: breast [63-67], prostate [68-71], imaging mass lesions of excised human liver [72].

## 3.3. Summary

In this chapter the most common techniques used for the elastography imaging are covered, amongst which the supersonic shear imaging and acoustic radiation force impulse are the most important techniques nowadays.

# Chapter 4 | Proposed Method

## 4.1.    Introduction

This chapter discusses in details the FEM construction, the ARFI generation and the acquisition sequence for tracking the wave propagation. A Finite Element Model (FEM) is constructed using the mechanical properties of the agar-gelatin phantom (by FEM software LISA FEA V8.0). To estimate the mechanical properties of this phantom by ultrasound imaging of the shear wave, a point force is applied for about 0.01 seconds in a transient manner. This point force is generated by FOCUS, a Matlab toolbox for generating the ARFI of specified transducer parameters. The resulting shear wave is then tracked by B-mode ultrasound imaging at a high frame rate and its speed is estimated to calculate the stiffness of the phantom; a mechanical property of the phantom; using Matlab software version R2010a. Tracking the shear wave propagation off-axis and measuring its velocity is a necessary process to have a quantitative map of elasticity for the tissue under consideration.

Tracking process is performed at several stationary nodes inside the model (probed nodes) along the lateral direction; where every node gives a curve of its displacement versus time profile from which we estimate the speed of the propagated wave. During its propagation, the shear wave covers a few meters per second, and a specific frame rate (FR) is needed to appropriately catch the propagating peak leading to a good estimation of the waves speed. Hence, a frame rate of several kilohertz is needed to have a good estimation for the speed since the conventional ultrasound scanners are not efficient due to their low frame rates; typically reaching about 50 to 60 frames per second [20].

## 4.2.    Methodology

The block diagram of the proposed method is shown in Fig. 4.1. The FEM is constructed so that it simulates the mechanical properties of the agar-gelatin phantom. Then a reference frame is acquired for the plane at which the point force is applied. After that an acoustic radiation force impulse is sent to that plane for which the shear wave speed is being calculated. A shear wave is formed and propagates off-axis to both lateral edges of the phantom. During its propagation deformed frames are acquired for recording its progress in the phantom. Comparing these deformed with respect to the reference frame gives rise to the propagating wave speed. Lateral Time-To-Peak displacement curves are used to estimate the wave speed. Hence an estimate for the mechanical properties of the investigated area is obtained.

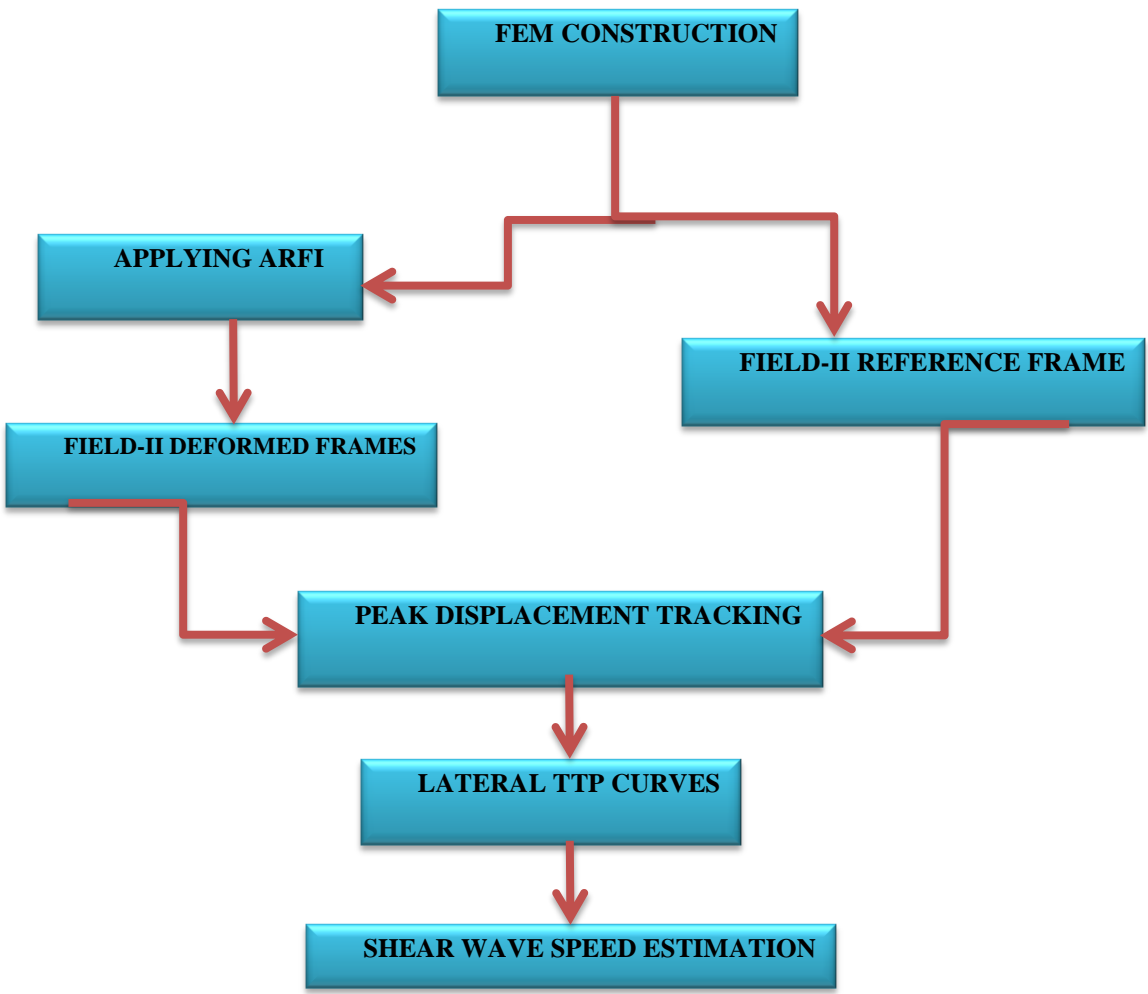

Figure 4.1: Block diagram of the proposed method.

### 4.2.1. FEM Mesh Construction

The experiment is performed using a viscoelastic agar-gelatin model. The choice of this material is due to the ability of the gelatin to maintain the stiffness of the phantom, and the ability of the agar to act as scatterers for the ultrasound waves in the phantom. A FEM mesh for this phantom is generated by Finite Element Modeling (FEM) software LISA FEA V8.0 to simulate its behavior when applying the Acoustic Radiation Force Impulse (ARFI). The FEM is a square shaped plate and unity in dimensions (1 meter side length), this square shaped plate resembles the plane inside which the ARFI is generated and the shear wave propagation takes place. Dividing this model into small squares; by Meshing techniques; provides an accurate displacement calculations' leading to a good estimation of the shear wave speed. Each sub-square is attached to its neighbourhood by nodes. The wave tracking is achieved by tracking the peak displacements using B-mode ultrasound imaging. At these nodes; differences in times for reaching the peak displacement at two or more successive nodes are calculated to provide an estimate for the speed. The mesh consists of:

- 1105 nodes and 1024 elements.
- The opposing face to the transducer is constrained completely.
- The face where the transducer touches is allowed to move in the perpendicular direction; i.e. the direction of the shear wave and other faces are allowed to move freely in all directions.

The distance between any two successive nodes in the lateral direction is 0.015625 meters. Assigning the material mechanical parameters of the agar-gelatin phantom for the simulated model to act as a real material, those properties are listed in Table 4.1 [42]. Figure 4.2 shows the phantom and the laterally probed nodes. The node at which the ARFI is applied is highlighted by the green arrow in the same figure. This design is the standard design to be used throughout the whole research and for any other mechanical parameters' phantoms.

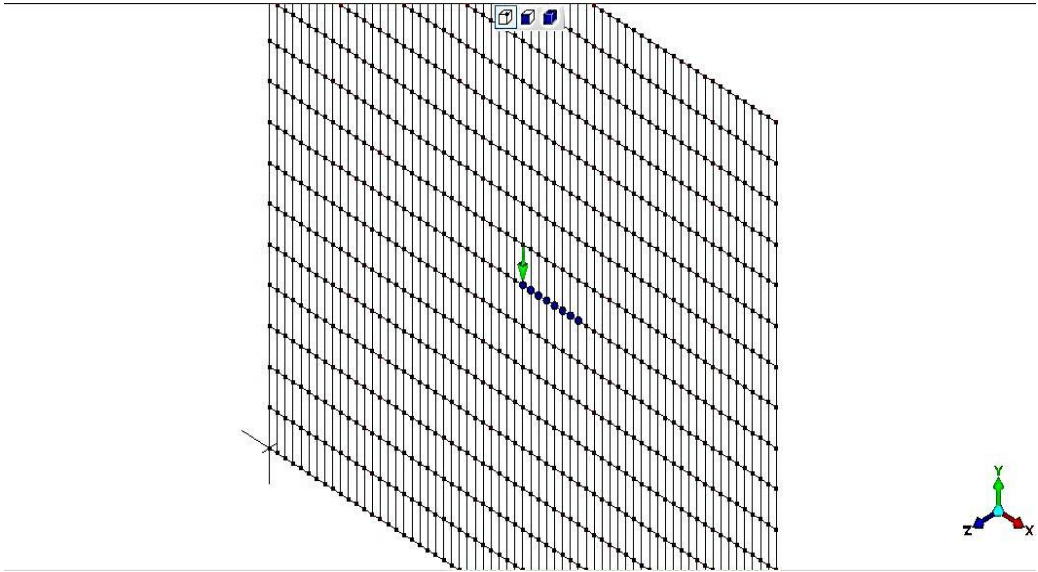

Figure 4.2: Phantom model with lateral probed nodes.

Table 4.1: Material mechanical parameters for agar-gelatin phantom [42].

| Material parameter | Value | Unit |
|---|---|---|
| Young's modulus ($E_o$) | 107585 | Pa |
| Shear modulus ($G_o$) | 35886 | Pa |
| Poisson's ratio ($\upsilon_o$) | 0.499 | --- |
| Density ($\rho$) | 1060 | $Kg/m^3$ |

## 4.2.2. Acoustic Radiation Force Impulse (ARFI) generation

Local displacements and shear waves are induced within the tissue due to the Acoustic Radiation Force (ARF) generated by the ultrasound transducer. Pressure field from such a transducer has to be simulated to determine the ARF field. Linear array ultrasound transducer model has been constructed using Focus toolbox in Matlab (an ultrasound tool for simulating the transducers and to calculate the resulting pressure field for a given material). The transducer model has been given the parameters listed in Table 4.2.

Table 4.2: Ultrasound transducer parameters.

| Parameter | Value |
|---|---|
| Number of elements | 128 |
| Elements' width | 170 µm |
| Kerf | 30 µm |
| Transducer width | 0.0256 m |
| Focal depth | 50 mm |
| Center frequency ($f_o$) | 12 MHz |

Apodization of a sinusoidal function is used when exciting the piezoelectric elements (**PE**). The edge elements are excited long before the central elements are excited in correspondence with the sine shape. Fast Nearfield Method (FNM) in Focus is used to calculate the pressure field in the plane of excitation. The resulting pressure distribution in the 2D field is shown in Fig. 4.3. The grating lobes are visible and clear at the distal edges of the phantom. The 2D pressure field at different axial positions is calculated and the pressure field at the focal plane is used in the FEM simulation. The FEM behavior to this focal 2D pressure field at the focal plane is investigated, and the resulting shear wave is tracked. The focal plane is chosen to be at 50 mm in the axial direction. The pressure distribution along the axial line passing through the focal point is calculated as well, and is shown in Fig 4.4. The axial distribution of the pressure shows that the highest pressure value is found around the focal zone at high frequencies.

The highest pressure value in the axial direction approaches the transducer surface at lower frequencies and moves forth away from the transducer surface as the frequencies increases. Also the axial beam intensity profile is calculated and shown in Fig. 4.5. The Acoustic Radiation Force Impulse (ARFI) is calculated along the axial direction and shown in Fig. 4.6. This acoustic radiation force impulse is used to excite the tissue and to induce the local displacement, resulting in shear wave propagation. This force is applied for a very short duration, typically 0.02 second and less. The force is applied in a gradually increasing manner, and is removed in a gradually decreasing manner as well; each part of the applying and the removal of the force take typically 0.01 second; as shown clearly in Fig. 4.7.

The time of investigation is 50 milliseconds; the force is induced within 10 milliseconds to reach its peak, and another 10 milliseconds to be fully removed and to reach zero magnitude, summing up for 20 milliseconds of excitation. The rest of the time is for the tracking B-mode imaging procedure.

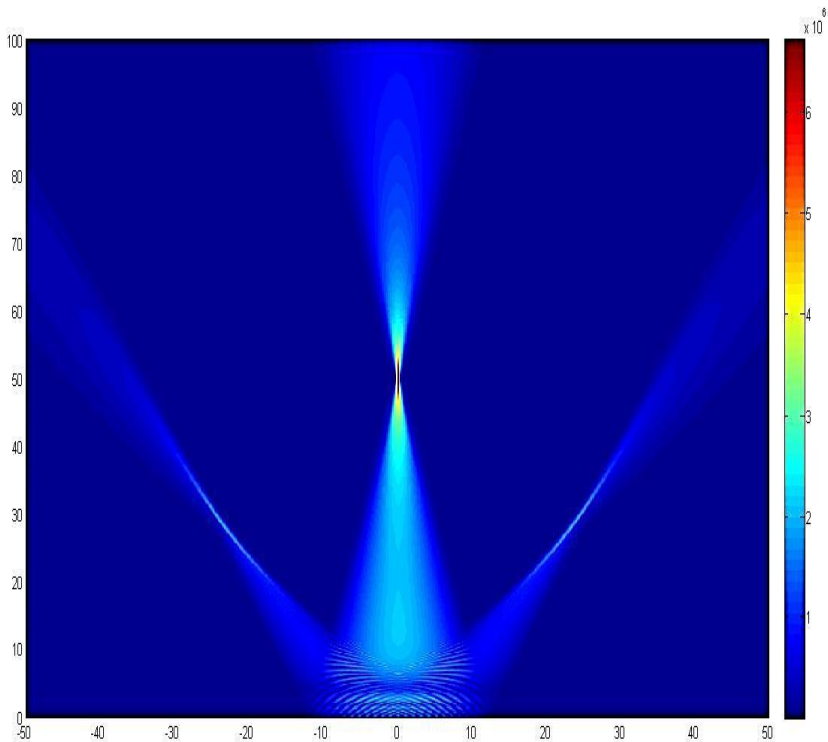

Figure 4.3: Two dimensional pressure field.

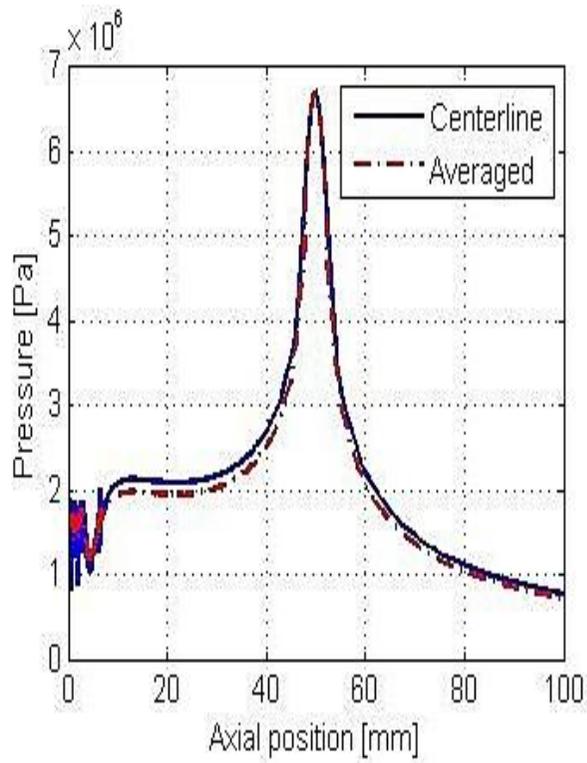

Figure 4.4: Pressure distribution along the axial line passing through the focal point.

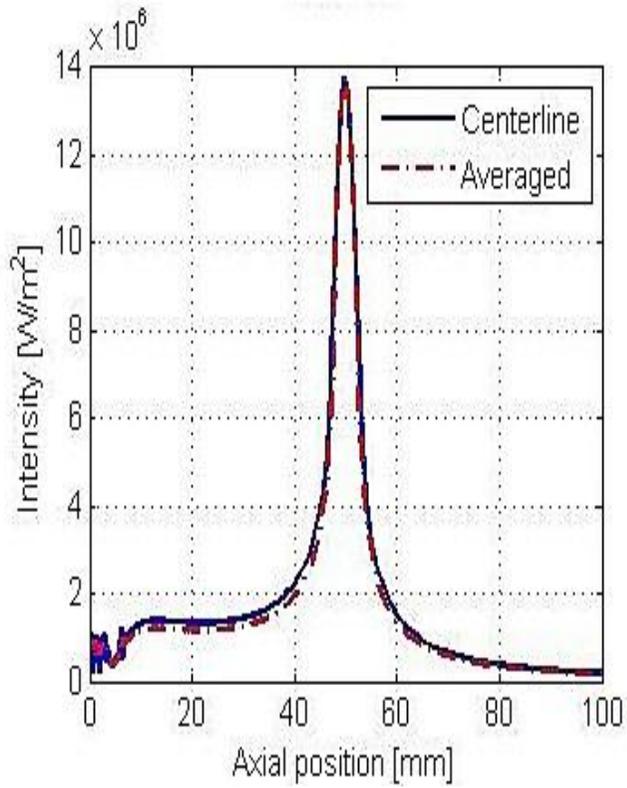

Figure 4.5: Axial beam intensity profile.

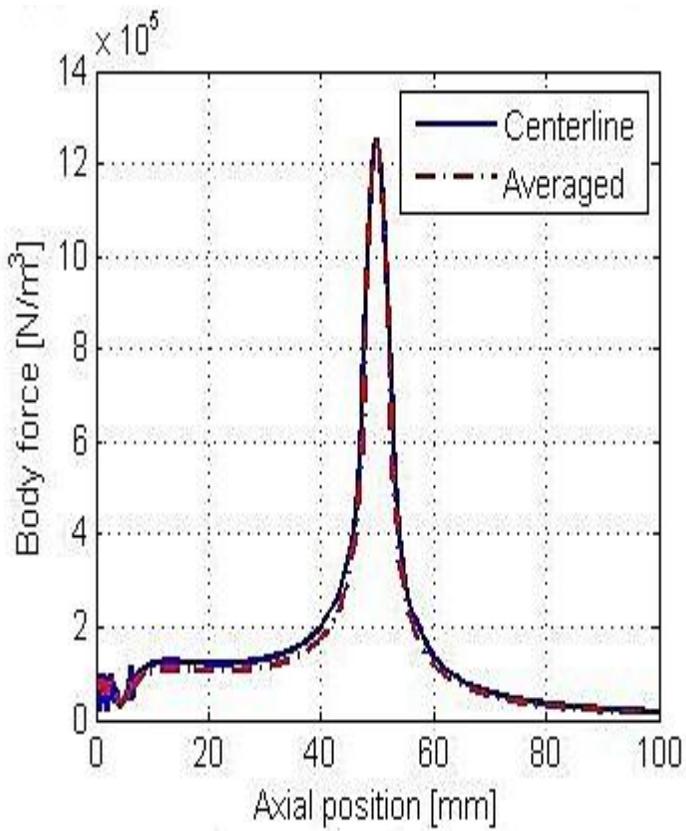

Figure 4.6: Axial Acoustic Radiation Force Impulse.

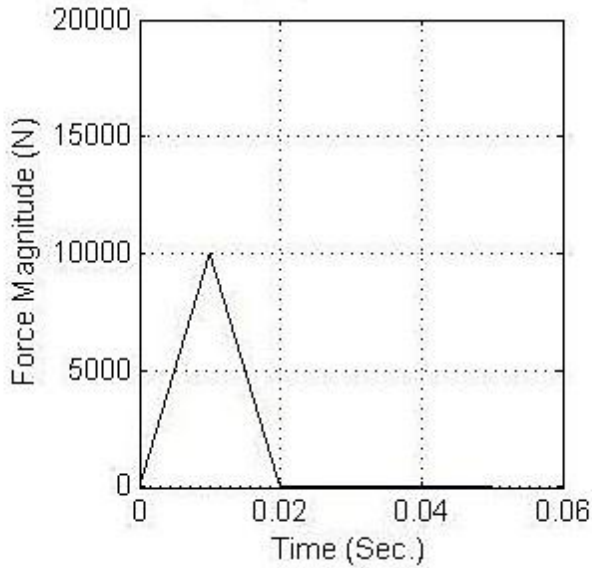

Figure 4.7: Force magnitude versus the time of excitation.

### 4.2.3. Acquisition Sequence

At the beginning, the medium is investigated by using a plane wave. This results in providing a reference frame. Then, the pushing sequence is sent by focusing the ultrasound beam to the area of interest (the focal zone of the beam is the area of interest) for a very short duration, typically about 20 milliseconds.

Just after the generation of the pushing pulse, an ordinary B-mode imaging procedure is carried out to catch the progress of the wave propagation through the model [20]. The B-mode imaging of the shear wave propagation is carried out by a very high frame rate, almost 1000 frames per second and up [20]. For investigating any other region in the model, the last procedure is repeated after modifying the focal zone of the beam.

To visualize the wave propagation, the peak of the wave is plotted versus time profile for the whole model. The speed can be estimated from this visualization. Converting the 3D visualizations to 2D curves of nodes displacements versus time profiles is more efficient for estimating the speed of the shear wave as proposed in [33]. For more feasibility of the calculations, one of the sides of the curve is selected to calculate the shear wave speed by tracking the peak displacement. This was proposed in [73] by Ned C. Rouze *et al*.

### 4.2.4. Shear Wave Speed Estimation

The shear wave speed is estimated using the Time-To-Peak displacement (TTP) method. Time-To-Peak displacement is defined as the time taken by a specific part of tissue to reach its maximum displacement. It is a characteristic for each tissue. Simply, the shear wave speed is estimated as the distance difference between two nodes divided by the time difference at which the TTPs are occurring for these two nodes. Yet, for a good estimation of the shear wave speed, more nodes are involved in the calculation where the average is calculated.

For calculating the TTP, the displacement profile for each node with time must be obtained. In other words, to observe the specific time at which the node reaches its peak displacement. The displacement profile for the central node of excitation is obtained as shown in Fig. 4.8. It is clear that the maximum displacement takes place at t = 0.014 sec. given that the force is applied at the central node.

Having displacement profiles for the nodes that are laterally away from the node at which the excitation happens, allows precise estimation of the shear wave speed, by calculating their successive TTPs.
Figure 4.9 shows clearly displacements profiles for five successive nodes away from the central node used to calculate the TTPs. Moving outwards from the central node, peaks are gradually decreased. The shear wave speed is calculated as the average value of distance difference between two successive nodes divided by the difference in the TTPs for these same two nodes, for eight nodes, and given by:

$$C_{n,n-1} = \frac{\Delta x}{\Delta t} \qquad\qquad (4.1)$$

$$C_{avg} = \frac{\sum_{i=1}^{n} C_{n,n-1}}{n} \qquad\qquad (4.2)$$

## 4.3.    Results

In this section, results for the agar-gelatin phantom are reported using the mechanical properties presented in Table 4.1. Results are obtained by using both the Finite Element Analysis software LISA in conjunction with Matlab software.

Tables 4.3, 4.4, and 4.5 shows different values obtained for the shear wave speeds at each successive two nodes. The average speed is found to be 5.2083 (m/sec) at its optimum case of 0.001 seconds as a time difference between each two successive frames.

The average is taken for eight different nodes, but for the feasibility of displaying the curves and observing them, only displacements for five nodes are shown

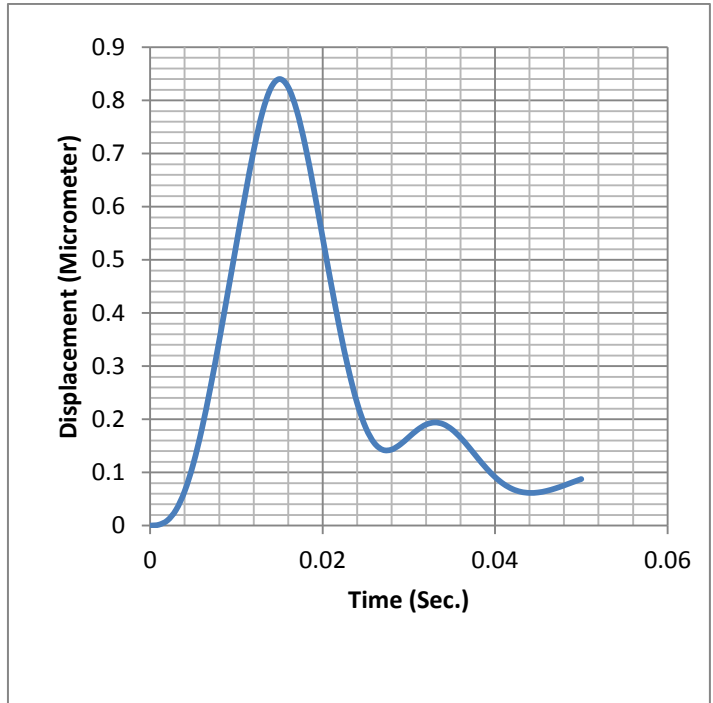

Figure 4.8: Displacement magnitude profile at the central node.

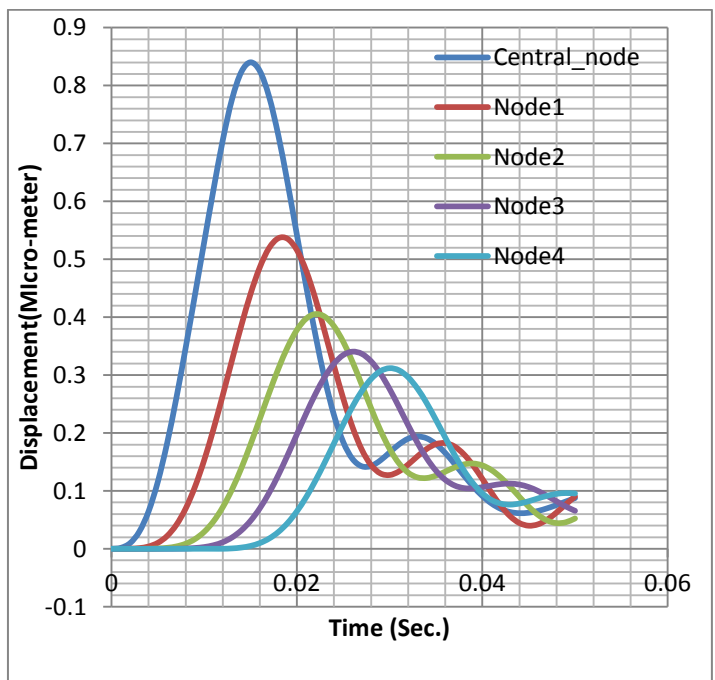

Figure 4.9: Displacements profiles for five successive nodes away from the central node.

Table 4.3: Shear wave speeds at 8 different nodes at a frame rate of 1 KHz.

| Node no. | $\Delta x$ (m) | $\Delta t$ (sec.) | C (m/sec.) |
|----------|----------------|-------------------|------------|
| 0, 1 | 0.015625 | 0.003 | 5.2083 |
| 1, 2 | 0.015625 | 0.003 | 5.2083 |
| 2, 3 | 0.015625 | 0.003 | 5.2083 |
| 3, 4 | 0.015625 | 0.003 | 5.2083 |
| 4, 5 | 0.015625 | 0.003 | 5.2083 |
| 5, 6 | 0.015625 | 0.003 | 5.2083 |
| 6, 7 | 0.015625 | 0.003 | 5.2083 |

Table 4.4: Shear wave speeds at 8 different nodes at at a frame rate of 1.1 KHz.

| Node no. | $\Delta x$ (m) | $\Delta t$ (sec.) | C (m/sec.) |
|----------|----------------|-------------------|------------|
| 0, 1 | 0.015625 | 0.003 | 5.2083 |
| 1, 2 | 0.015625 | 0.004 | 3.9063 |
| 2, 3 | 0.015625 | 0.004 | 3.9063 |
| 3, 4 | 0.015625 | 0.004 | 3.9063 |
| 4, 5 | 0.015625 | 0.004 | 3.9063 |
| 5, 6 | 0.015625 | 0.003 | 5.2083 |
| 6, 7 | 0.015625 | 0.003 | 5.2083 |

Table 4.5: Shear wave speeds at 8 different nodes at a frame rate of 2 KHz.

| Node no. | $\Delta x$ (m) | $\Delta t$ (sec.) | C (m/sec.) |
|----------|----------------|-------------------|------------|
| 0, 1 | 0.015625 | 0.006 | 2.6042 |
| 1, 2 | 0.015625 | 0.007 | 2.2321 |
| 2, 3 | 0.015625 | 0.007 | 5.2083 |
| 3, 4 | 0.015625 | 0.003 | 5.2083 |
| 4, 5 | 0.015625 | 0 | NaN |
| 5, 6 | 0.015625 | 0 | NaN |
| 6, 7 | 0.015625 | 0 | NaN |

## 4.4.    Discussion

From previous tables, estimating the shear wave speed does not give a fixed value for the speed. On other words, there is a tradeoff between the estimated speed and the frame rate used for the estimation process. There is an optimum frame rate that leads to the estimation of the closest speed value, although higher frame rates give closest estimated value.

In our experiments, the optimum frame rate is found to be 1 KHz (each frame of imaging takes about 0.001 sec). Other frame rates were found to give fluctuating velocities around 5 m/seconds as shown in Fig. 4.10. The optimum frame rate in these experiments led to a velocity of 5.2 m/sec. This is a characteristic for this phantom.

In the first laterally probed velocities curve (Fig. 4.10.a), it is observed that velocities are fixed and independent on the node lateral position inside the phantom. Other laterally probed velocities curves show the instability of the speed and its dependency on the node lateral position.

The calculated shear wave speed for the phantom under study is about 5.7 m/sec. The deviation of our results from the calculated is due to the insufficient number of nodes involved in the phantom construction, yet the speed estimation does not give a precise value as the calculated one. Moreover, the distance between the nodes in the x-direction is predicted to have a role in a good estimation of the speed. As the nodes are the stationary stations from which we monitor the propagation of the wave, less distance means better tracking for the peak displacement.

All methods using external actuators and external vibrating sources are introducing a proper deformation value, which is large enough to be picked up and processed. This facilitates the measurement process of tissue nonlinearities. This also has the drawback of complex hardware to achieve such a function. Amongst these methods; sonoelasticity and transient methods.

On the other hand, methods which utilize ARFI have the advantage of using the same ultrasound scanner for excitation and imaging, i.e. the same scanner for probing and measurement. The only disadvantage, here, is that the intensity of the excitation pulse must be kept under certain limit due to the mechanical and thermal considerations when dealing with ultrasound waves [74].

These unwanted bioeffects limits the intensity of excitation, and hence the induced deformation to less than 30 μm. This also limits the shear wave's estimation beyond 6 or 7 cm, due to high attenuation. Furthermore, as mentioned earlier, the high attenuation is considered an advantage, where it produces highly localized shear waves [74].

As far as we know, there is no formula to predict the relationship between the excitation frequency and the elasticity moduli. If it is taken into consideration that the tissue particles may be modeled as a vibrating pendulum, there will be some excitation frequency that will have no action on these particles, as their moment of inertia will be very high for that frequency.

In other words, shear waves will not be generated and hence the tissue will be misestimated to be having low modulus of elasticity, but, in reality it is not. This critical frequency or any frequencies approaching it should not be used as they deflect the estimation. This critical excitation frequency will be investigated in the future work.

## 4.5. Summary

In this chapter, shear wave speed is estimated for the Agar-gelatine. A phantom with specific mechanical properties from literature is used. A model for this phantom is generated by a finite element modeling software to simulate its behavior. A point source force is applied at the focal point, which is the center of the phantom, to stimulate the shear wave propagation. TTP method is used to estimate the speed. The estimated speed is found to be 5.2 m/sec; while the calculated one is 5.7 m/sec. The difference between both of them arises from the shortage of the number of nodes by which the model has been constructed.

In this chapter, a finite element model is constructed so that is simulates the mechanical properties of the agar-gelatin phantom using the finite element analysis software LISA V8.0. The acoustic radiation force impulse is generated using FOCUS, the matlab tool, to simulate the acoustic force to be applied on the phantom to excite the shear wave. The shear wave propagation acquisition procedure is also introduced in this chapter. The estimation is performed at eight different nodes along the lateral direction of the phantom on either sides. The speed is simply the average of the speeds at these successive nodes. Finally the results are shown to be around 5 m/sec. which is a proper estimation for the shear wave speed at the optimum frame rate of 1 KHz.

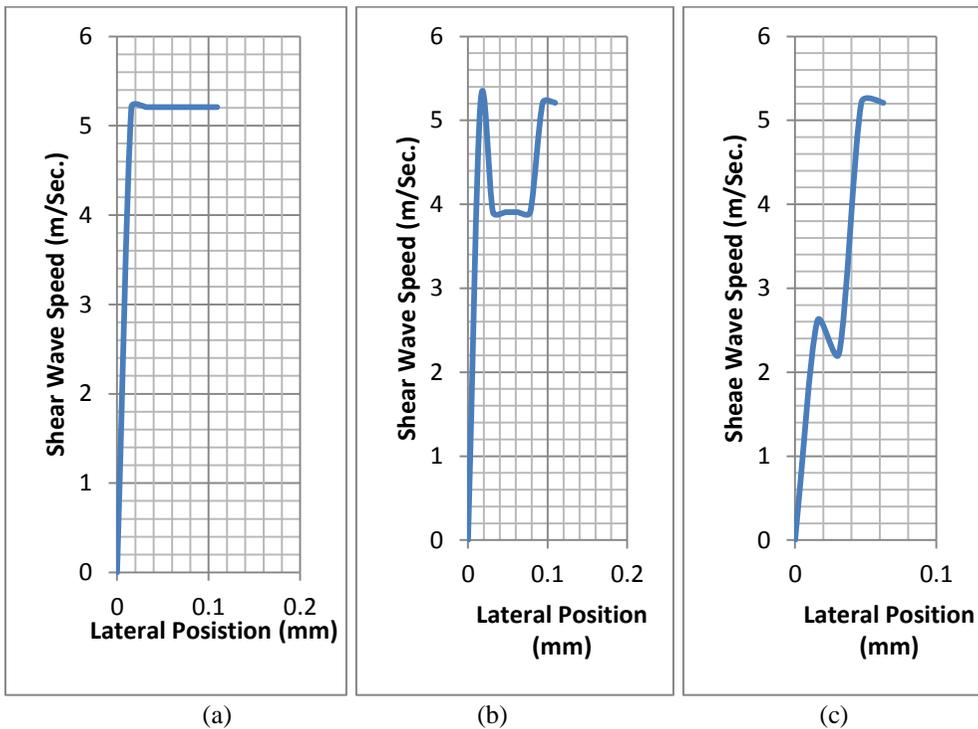

(a)                    (b)                    (c)

Figure 4.10: Laterally probed velocities curves for different frame rates: (a) 1KHz, (b) 1.1KHz and (c) 2KHz

# Chapter 5 | Case Studies

## 5.1. Introduction

In this chapter, experiments of five different Young's moduli phantoms and case studies of two different soft tissues types are introduced. Breast tissue is one of them, liver tissue is the other.

## 5.2. Preliminary Experiments

Calculating the shear wave speed for five phantoms with different elastic moduli is performed as well. These five different moduli are listed in Table 5.1 [73].

Table 5.1: Elastic moduli values of five different phantoms.

| Phantom | Young's Modulus |
|---------|-----------------|
| Phantom 1 | 5.2 KPa |
| Phantom 2 | 9.8 KPa |
| Phantom 3 | 23.9 KPa |
| Phantom 4 | 44.2 KPa |
| Phantom 5 | 67.3 KPa |

These phantoms are all agar-gelatin phantoms but with different Young's moduli. These phantoms are set to have the same density, typically 1060 $Kg/m^3$, and the same Poisson's ratio, typically 0.499.

## 5.2.1. Methodology

The same method discussed in details in chapter 4 is used to estimate the shear wave speed for these five phantoms. The same ultrasound transducer parameters that are used with the 107.585 KPa phantom, are used with these five phantoms. The same ARFI used for exciting the 107.585 KPa phantom is used to excite these five phantoms as shown in Fig. 4.6. The same durations for excitation and B-mode imaging procedure are also used.

## 5.2.2. Results

Results for shear wave propagation for these five phantoms are shown in Fig. 5.1. Resulting curves for shear wave propagation for different elastic moduli five phantoms where; the left hand side curves illustrate the displacement magnitude profile for the central node, the middle curves illustrate the displacement magnitudes at five probed nodes and the right hand side curves illustrate the velocity curve versus the laterally probed nodes; for: a) 5.2 KPa, b) 9.8 KPa, c) 23.9 KPa, d) 44.2 KPa, e) 67.3 KPa. The average estimated speeds for corresponding shear waves are listed in Table 5.2. Average is calculated for eight different stationary nodes (probed nodes) along the lateral direction at the focal plane.

Table 5.2: Estimated and Calculated Shear Wave Speed for five different phantoms.

| Phantom | Average Shear Wave Speed (m/sec.) (Estimated) | Average Shear Wave Speed (m/sec.) (Calculated) | Average Shear Wave Speed (m/sec.) (Radon transform [73] ) | Accuracy % of estimated speed w.r.t the calculated speed |
|---------|---------|---------|---------|---------|
| Phantom 1 | 1.1347 | 1.32 | 1.53 | 85.6% |
| Phantom 2 | 1.4689 | 1.81 | 2.19 | 80.6% |
| Phantom 3 | 2.2853 | 2.82 | 3.38 | 80.6% |
| Phantom 4 | 3.1622 | 3.84 | 4.47 | 82.2% |
| Phantom 5 | 3.8690 | 4.74 | 5.48 | 81.43% |

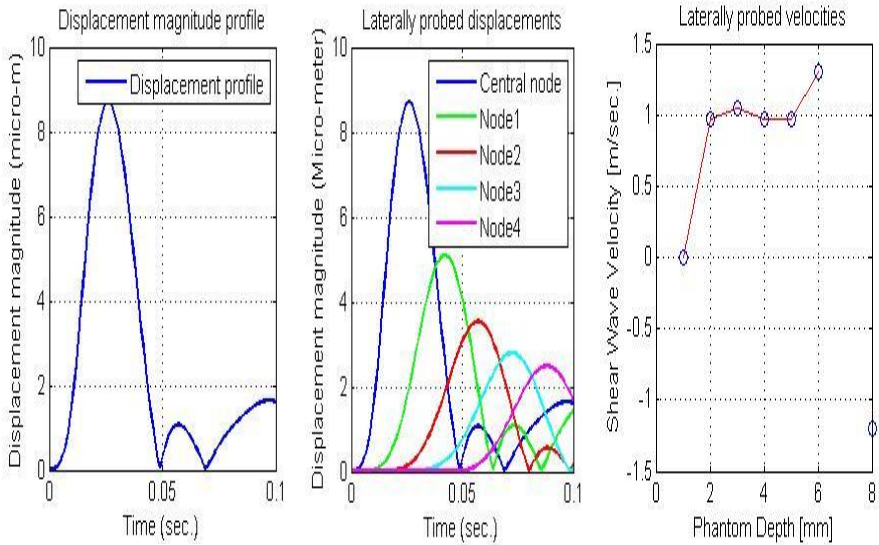

(a)

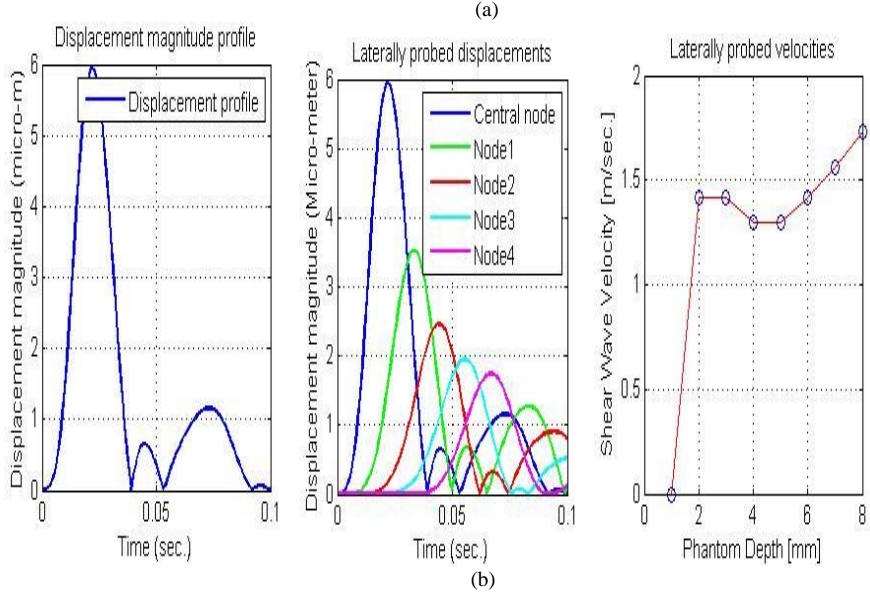

(b)

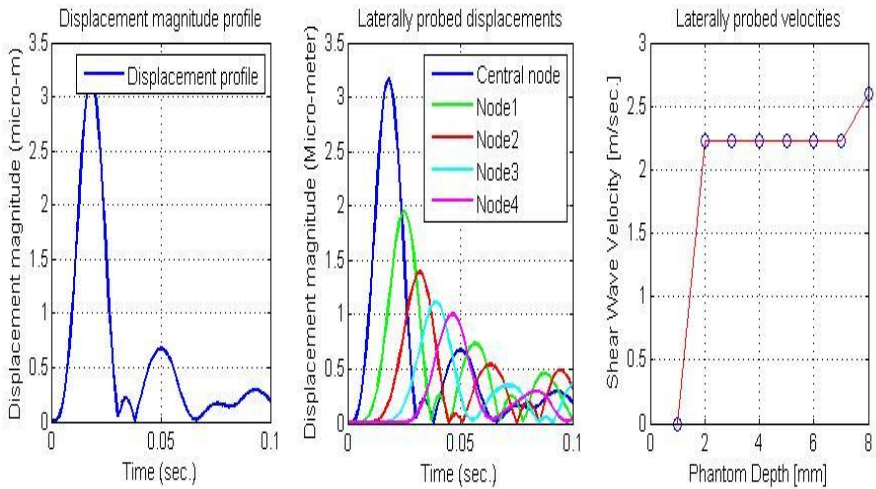

(c)

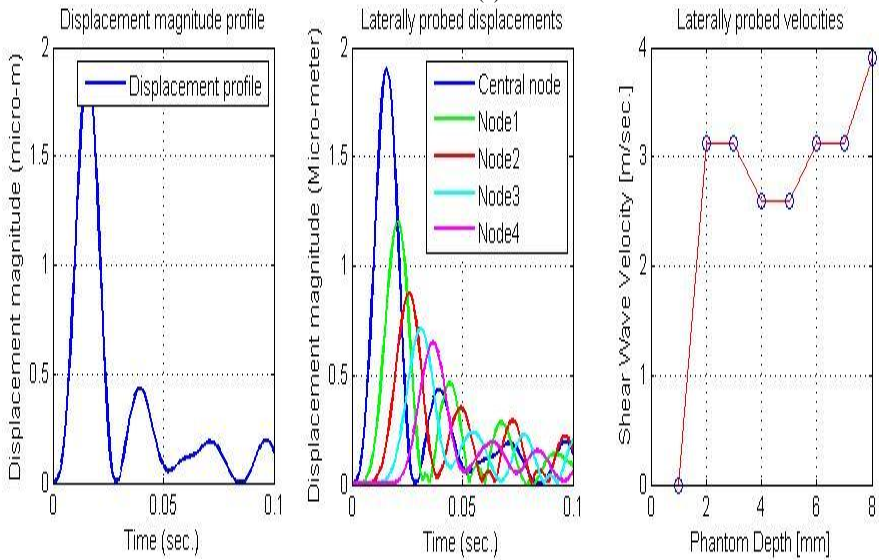

(d)

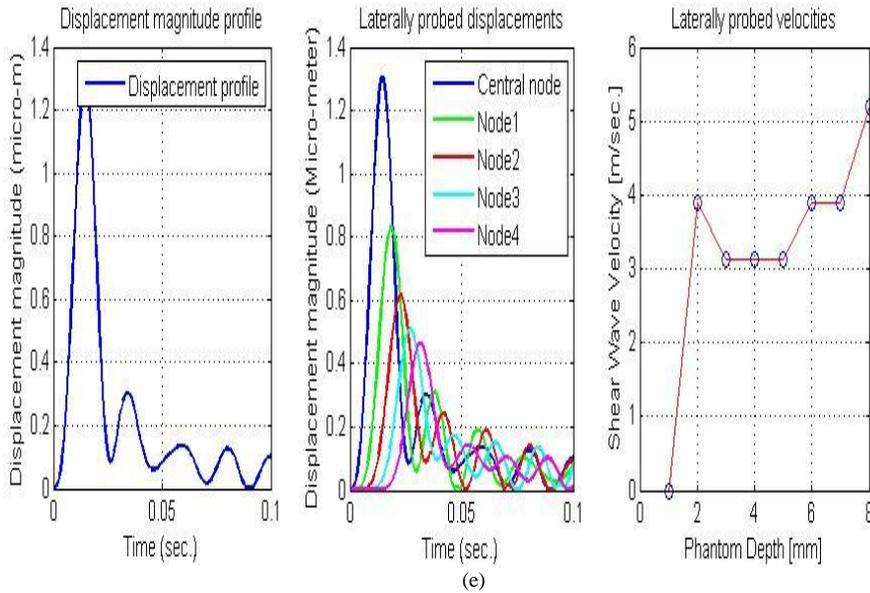

(e)

Figure 5.1: Resulting curves for shear wave propagation for different elastic moduli five phantoms where; the left hand side curves illustrate the displacement magnitude profile, the middle curves illustrate the displacement magnitudes at five probed nodes and the right hand side curves illustrate the velocity curve versus the laterally probed nodes; for: a) 5.2 KPa, b) 9.8 KPa, c) 23.9 KPa, d) 44.2 KPa, e) 67.3 KPa.

These results for shear wave speed estimation are all obtained at the optimum frame rate at which the shear wave speed of 107.585 KPa phantom has been obtained, typically 1 KHz, as discussed in chapter 4.

Results obtained from Radon transform method [73] and the accuracy of the obtained results with respect to the calculated speeds are listed in Table 5.2 as well. The estimated speeds results are 1.13, 1.46, 2.28, 3.16 and 3.86 (m/sec.) compared to the calculated value 1.32, 1.81, 2.82, 3.84 and 4.74 (m/sec.) for 5.2, 9.8, 23.9, 44.2 and 67.3 KPa phantoms respectively.

### 5.2.3.  Discussion

The highest accuracy value is observed at 5.2 KPa phantom and the lowest one is observed at both 9.8 and 23.9 KPa phantoms. Yet the most stable laterally probed velocities curve is obtained at 23.9 KPa, to give a shear wave speed of 2.28 (m/sec.).

In order, the 9.8 KPa phantom show a relative stability in the speed estimation with respect to the lateral position. The 5.2 KPa phantom holds the $3^{rd}$ place in the issue of estimation stability. After that, the 44.2 KPa and 67.3 KPa hold the last place with unstable laterally probed velocities. The stability of the estimated speed gives rise to the independency of the estimation on the lateral position.

Also, the second highest peak is observed to occur delayed in Fig. 5.1.a, this is coincident with the fact of having a lower shear wave speed for lower stiffness phantoms. As the phantom stiffness increases, the time at which this peak occurs approaches the time mark of 0.05 seconds. It is at 23.9KPa where this peak occurs typically at 0.05 seconds. Higher stiffness picks this peak to the left of the time mark of 0.05 seconds.

### 5.2.4. Conclusion

There is a relationship between the time of occurrence of this peak (the $2^{nd}$ highest peak in decreasing order) and the elastic modulus of the phantom model, given the condition that the information are extracted from the central node displacement magnitude profile.

### 5.3. Case Studies

Breast tissue is studied as normal, normal glandular and fibrous glandular cases. Liver tissue is investigated as normal and cirrhosis cases. Normally, stiffer tissues tend to have a higher young's modulus than normal soft tissues.

**Breast**

The breasts have a conical shape and are located, one on each side, within the subcutaneous layer of the thoracic wall, anteriorly to the pectoralis major muscle. They extend superiorly as far as the level of the second rib, inferiorly as far as the level of the sixth or seventh ribs, laterally as far as the anterior axillary line (sometimes as far as the middle axillary line) and medially they reach the lateral margin of the sternum. Posteriorly, they make contact with the fascia of the pectoralis major, serratus anterior and obliquus externus muscles and the most cranial portion of the rectus abdominis muscle. Its base is circular and measures around 10 to 12 cm, but its volume is very variable. The weight of a non-lactating breast ranges from 150 to 225g, while a lactating breast may exceed 500g in weight. The breasts of nulliparous women have a hemispherical shape, while those of multiparous women are broader and pendulant. With aging, the breast volume decreases, and the breast becomes less firm, flatter and pendulant. Three portions are distinguished anatomically: the gland itself (glandula mammaria), the mammary papilla (papilla mammariae) and the areola (areola mammae). The mammary gland is formed by fifteen to twenty lobes (lobi glandulae mammariae) that are arranged radially and delimited by septa of conjunctive tissue and adipose tissue in the subcutaneous layer as shown in Fig.5.2. The mammary parenchyma is more abundant in the upper half of the gland, especially in the superolateral quadrant. The mammary tissue frequently extends beyond the apparent outline of the breast, projecting towards the axilla as an axillary process (sometimes called tail of Spence). The principal duct of each lobe, the lactiferous duct (ductus lactiferi), opens separately into the mammary papilla. In turn, the lobe is formed by smaller functional units, the lobules (lobuli), from which ducts converge towards the main duct of the lobe. The subcutaneous layer (tela subcutanea) completely surrounds the gland, except in the region of the papilla. It needs to be clarified that the subcutaneous layer was in the past named the superficial fascia. The part of this layer located immediately in front of the fascia of the pectoralis major muscle was erroneously called the deep layer of the superficial fascia. In the subcutaneous layer, fascicles of conjunctive tissue are observed to permeate the lobes and lobules, particularly in the upper part of the gland, which cross the breast anteroposteriorly, extending from the dermis to the part of the subcutaneous layer next to the fascia of the pectoralis major muscle. These fascicles are known as the suspensory (or Cooper's) ligaments of the breast (suspensoria mammaria ligament). Neoplasms of the breast may affect them and cause localized retraction of the overlying skin. Although not officially recognized by the present anatomical terminology, the space located between the deep part of the subcutaneous layer and the muscular fascia of the pectoralis major muscle is known as the retromammary bursa, submammary serous bursa or also Chassaignac's bursa. It is easily identified during mastectomy. This space contributes towards the mobility of the breast on the thoracic wall. The mammary papilla represents the apex of the cone and contains the opening for all the lactiferous ducts from the lobes. Close to the apex of the papilla, each duct presents a distal saclike dilatation known as the lactiferous sinus (sinus lactiferi). It is worth emphasizing that, although the term nipple is habitually utilized in clinical practice, it is recommended that the expression mammary papilla should be utilized in the anatomical terminology. The areola is a slightly raised disc-shaped area of variable size surrounding the papilla. Initially, it has a rosy hue, but becomes irreversibly pigmented (chestnut brown) from the second month of gestation. On its surface, it presents granular and pointlike elevations known as areolar tubercles (tubercula areolares) or Montgomery's tubercles. These correspond to the anatomical representation of glands with intermediate histological structure between sudoriparous and mammary glands, the areolar glands (glandulae areolares) [75].

Fibrocystic breasts or fibrocystic breast disease or fibrocystic breast condition commonly referred to as "FBC" is a condition of breast tissue affecting an estimated 30-60% of women and at least 50% of women of childbearing age [76]. Some studies indicate that the lifetime prevalence of FBC may be as high as 70% to 90% [77-78]. It is characterized by noncancerous breast lumps which can sometimes cause discomfort, often periodically related to hormonal influences from the menstrual cycle[79].The changes in fibrocystic breast disease are characterised by the appearance of fibrous tissue and a lumpy, cobblestone texture in the breasts. These lumps are smooth with defined edges, and are usually free-moving in regard to adjacent structures. The bumps can sometimes be obscured by irregularities in the breast that are associated with the condition. The lumps are most often found in the upper, outer sections of the breast (nearest to the armpit), but can be found throughout the breast. Women with fibrocystic changes may experience a persistent or intermittent breast aching or breast tenderness related to periodic swelling. Breasts and nipples may be tender or itchy. Symptoms follow a periodic trend tied closely to the menstrual cycle. Symptoms tend to peak in the days and, in severe cases, weeks before each period and decrease afterwards. At peak, breasts may feel full, heavy, swollen, and tender to the touch. No complications related to breastfeeding have been found.

Diagnosis is mostly done based on symptoms after exclusion of breast cancer. Nipple fluid aspiration can be used to classify cyst type (and to some extent improve breast cancer risk prediction) but it is rarely used in practice. Biopsy or fine needle aspiration are rarely warranted [80].

Fibrocystic breast disease is primarily diagnosed based on the symptoms, clinical breast exam and on a physical exam. During this examination, the doctor checks for unusual areas in the breasts, both visually and manually. Also, the lymph nodes in the axilla area and lower neck are examined. A complete and accurate medical history is also helpful in diagnosing this condition. If the patient's medical history and physical exam findings are consistent with normal breast changes, no additional tests are considered but otherwise the patient will be asked to return a few weeks later for reassessment [81]. Women may detect lumps in their breasts during self-examination as well.

In order to establish whether the lump is a cyst or not, several imaging tests may be performed. Mammography is usually the first imaging test to be ordered when unusual breast changes have been detected during a physical examination. A diagnostic mammography consists in a series of x-rays that provide clear images of specific areas of the breast. Ultrasounds and MRIs are commonly performed in conjunction with mammography as they produce clear images of the breast and clearly distinguish between fluid-filled breast cysts and solid masses. The ultrasound and MRI exams can better evaluate dense tissue of the breast; hence it is often undergone by young patients, under 30 years old.

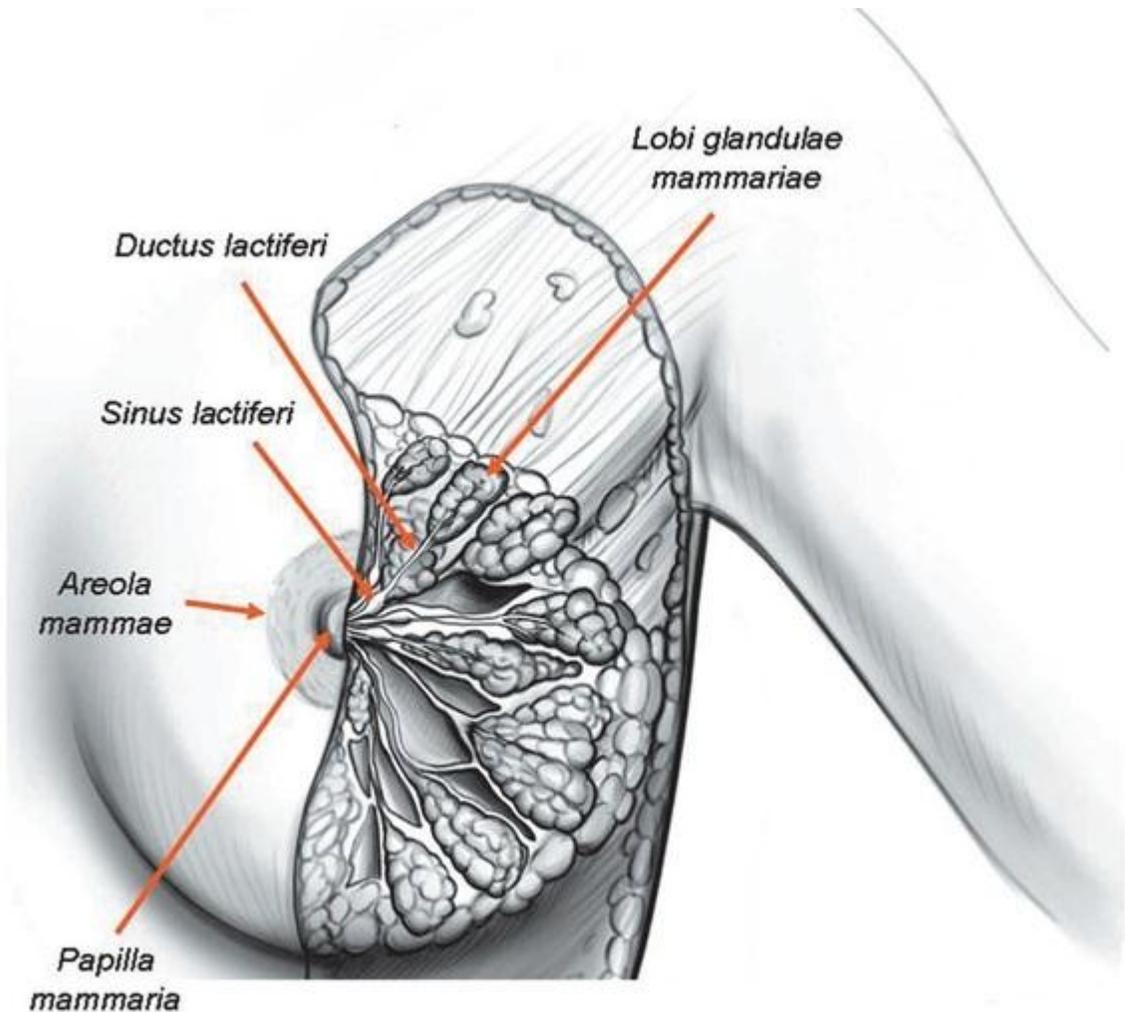

Figure 5.2 Right breast. The skin and subcutaneous layer have been extracted in order to view the lobi glandulae mammariae, ductus lactiferi and sinus lactiferi (anterior view) [75].

## Liver

The liver is a vital organ of vertebrates and some other animals [82]. In the human, it is located in the upper right quadrant of the abdomen, below the diaphragm. The liver has a wide range of functions, including detoxification of various metabolites, protein synthesis, and the production of biochemicals necessary for digestion [83]. The liver is a gland and plays a major role in metabolism with numerous functions in the human body, including regulation of glycogen storage, decomposition of red blood cells, plasma protein synthesis, hormone production, and detoxification [84]. It is an accessory digestive gland and produces bile, an alkaline compound which aids in digestion via the emulsification of lipids. The gallbladder, a small pouch that sits just under the liver, stores bile produced by the liver [85]. The liver's highly specialized tissue consisting of mostly hepatocytes regulates a wide variety of high-volume biochemical reactions, including the synthesis and breakdown of small and complex molecules, many of which are necessary for normal vital functions [86]. Estimates regarding the organ's total number of functions vary, but textbooks generally cite it being around 500 [85].

The liver is a reddish brown wedge-shaped organ with four lobes of unequal size and shape. A human liver normally weighs 1.44–1.66 kg (3.2–3.7 lb) [84]. It is both the heaviest internal organ and the largest gland in the human body. Located in the right upper quadrant of the abdominal cavity, it rests just below the diaphragm, to the right of the stomach and overlies the gallbladder [84]. The liver is connected to two large blood vessels: the hepatic artery and the portal vein as shown in Fig. 5.3. The hepatic artery carries oxygen-rich blood from the aorta, whereas the portal vein carries blood rich in digested nutrients from the entire gastrointestinal tract and also from the spleen and pancreas [86]. These blood vessels subdivide into small capillaries known as liver sinusoids, which then lead to a lobule. Lobules are the functional units of the liver. Each lobule is made up of millions of hepatic cells (hepatocytes) which are the basic metabolic cells. The lobules are held together by fine areolar tissue which extends into the structure of the liver, by accompanying the vessels (veins and arteries) ducts and nerves through the hepatic portal, as a fibrous capsule called **Glisson's capsule** [86]. The whole surface of the liver is covered in a serous coat derived from peritoneum and this has an inner fibrous coat (Glisson's capsule) to which it is firmly adhered. The fibrous coat is of areolar tissue and follows the vessels and ducts to support them.

Now we turn to liver cirrhosis, where Cirrhosis is a condition in which the liver does not function properly due to long-term damage. Typically, the disease comes on slowly over months or years. Early on, there are often no symptoms. As the disease worsens, a person may become tired, weak, itchy, have swelling in the lower legs, develop yellow skin, bruise easily, have fluid buildup in the abdomen, or develop spider-like blood vessels on the skin. The fluid build-up in the abdomen may become spontaneously infected. Other complications include hepatic encephalopathy, bleeding from dilated veins in the esophagus or dilated stomach veins, and liver cancer. Hepatic encephalopathy results in confusion and possibly unconsciousness [86].

Cirrhosis is most commonly caused by alcohol, hepatitis B, hepatitis C, and non-alcoholic fatty liver disease [86]. Typically, more than two or three drinks per day over a number of years is required for alcoholic cirrhosis to occur. Non-alcoholic fatty liver disease is due to a number of reasons, including being overweight, diabetes, high blood fats, and high blood pressure. A number of less common causes include autoimmune hepatitis, primary biliary cirrhosis, hemochromatosis, certain medications, and gallstones. Cirrhosis is characterized by the replacement of normal liver tissue by scar tissue. These changes lead to loss of liver function. Diagnosis is based on blood testing, medical imaging, and liver biopsy [86].

Cirrhosis has many possible manifestations. These signs and symptoms may be either as a direct result of the failure of liver cells or secondary to the resultant portal hypertension. There are also some manifestations whose causes are nonspecific, but may occur in cirrhosis. Likewise, the absence of any does not rule out the possibility of cirrhosis. Cirrhosis of the liver is slow and gradual in its development. It is usually well advanced before its symptoms are noticeable enough to cause alarm. Weakness and loss of weight may be early symptoms.

The gold standard for diagnosis of cirrhosis is a liver biopsy, through a percutaneous, transjugular, laparoscopic, or fine-needle approach. A biopsy is not necessary if the clinical, laboratory, and radiologic data suggests cirrhosis. Furthermore, there is a small but significant risk to liver biopsy, and cirrhosis itself predisposes for complications caused by liver biopsy. The best predictors of cirrhosis are ascites, platelet count <160,000/mm3, spider angiomata and Bonacini cirrhosis discriminant score greater than 7.

Ultrasound is routinely used in the evaluation of cirrhosis. It may show a small and nodular liver in advanced cirrhosis along with increased echogenicity with irregular appearing areas. Other findings suggestive of cirrhosis in imaging are an enlarged caudate lobe, widening of the liver fissures and enlargement of the spleen. An enlarged spleen (splenomegaly), which normally measures less than 11–12 cm in adults, is suggestive of cirrhosis with portal hypertension in the right clinical setting. Ultrasound may also screen for hepatocellular carcinoma, portal hypertension, and Budd-Chiari syndrome (by assessing flow in the hepatic vein).

Cirrhosis is diagnosed with a variety of elastography techniques. Because a cirrhotic liver is generally stiffer than a healthy one, imaging the liver's stiffness can give diagnostic information about the location and severity of cirrhosis. Techniques used include transient elastography, acoustic radiation force impulse imaging, supersonic shear imaging and magnetic resonance elastography. Compared to a biopsy, elastography can sample a much larger area and is painless. It shows reasonable correlation with the severity of cirrhosis [86].

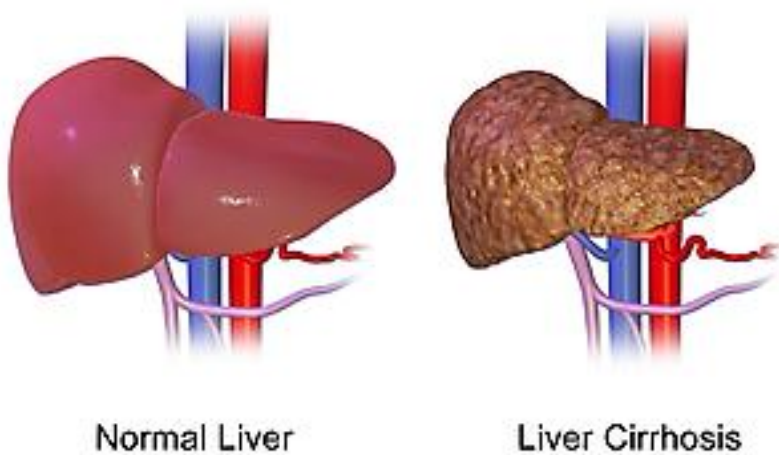

Figure 5.3: Normal liver and cirrhosis liver.

Typical values of Young's modulus for these tissues involved in case studies are listed in Table 5.3 [78-80].

Density of the soft tissue remains relatively constant in the body, very close to that of water ($10^3$ Kg/m$^3$). Average densities for these tissues are set to have a mean value $10^3$ (Kg/m$^3$) with tolerance up to 8% around the mean value of density. In the proceding experiments, the density is set to have $10^3$+6% (Kg/m$^3$).

Table 5.3: Young's moduli of elasticity for different soft tissues [78-80].

| Tissue type | Young's Modulus (in KPa) |
|---|---|
| Normal fatty breast tissue | 16-24 |
| Normal glandular breast tissue | 28-66 |
| Fibrous glandular breast tissue | 66-244 |
| Normal liver tissue | 0.4-6 |
| Cirrhosis liver tissue | 15-100 |

## 5.3.1. Method

Models for these tissues are made, and tested versus transient force impulse to induce a local displacement, and the resulting shear waves values' are estimated. The same method for generating the Mesh model; that is discussed in chapter 4 in details; is used to generate Mesh models for these tissues mentioned above in Table 5.3.

Also, the same transducer parameters used to excite the agar-gelatin phantom with ARF impulse in chapter 4 are used here in these case studies. The same acquisition sequence is used as well.

## 5.3.2. Results

Results for the shear wave speed estimation for these cases are estimated at two elastic moduli values, the upper value and the lower value for each tissue case. Normal fatty breast tissue for example, has got its induced shear wave speed estimated at the upper value of Young's modulus; typically at 24 KPa; and at the lower value of Young's modulus; typically at 16 KPa; resulting in a range of shear wave speed values that are expected while investigating this type of tissue in later clinical case studies.

**Breast Tissue**

Breast tissue cases have their resulting shear wave speeds shown in Fig. 5.4, Fig 5.5 and Fig. 5.6 respectively. Figure 5.4 shows the displacement magnitude profile for the central node and for five lateral nodes away from it and the laterally probed velocities for the normal fatty breast tissue. Figure 5.5 shows the displacement magnitude profile for the central node and for five lateral nodes away from it and the laterally probed velocities for the normal glandular breast tissue. Figure 5.6 shows the same curves but for the fibrous glandular breast tissue.

**Liver Tissue**

Results for liver tissue cases are shown in Fig. 5.7 and 5.8. Figure 5.7 shows the displacement magnitude profile for the central node and for five lateral nodes away from it and the laterally probed velocities for the normal liver tissue. Figure 5.8 shows the same curves but for the cirrhosis liver tissue. Shear wave speed estimated values for these models are listed in Table 5.4.

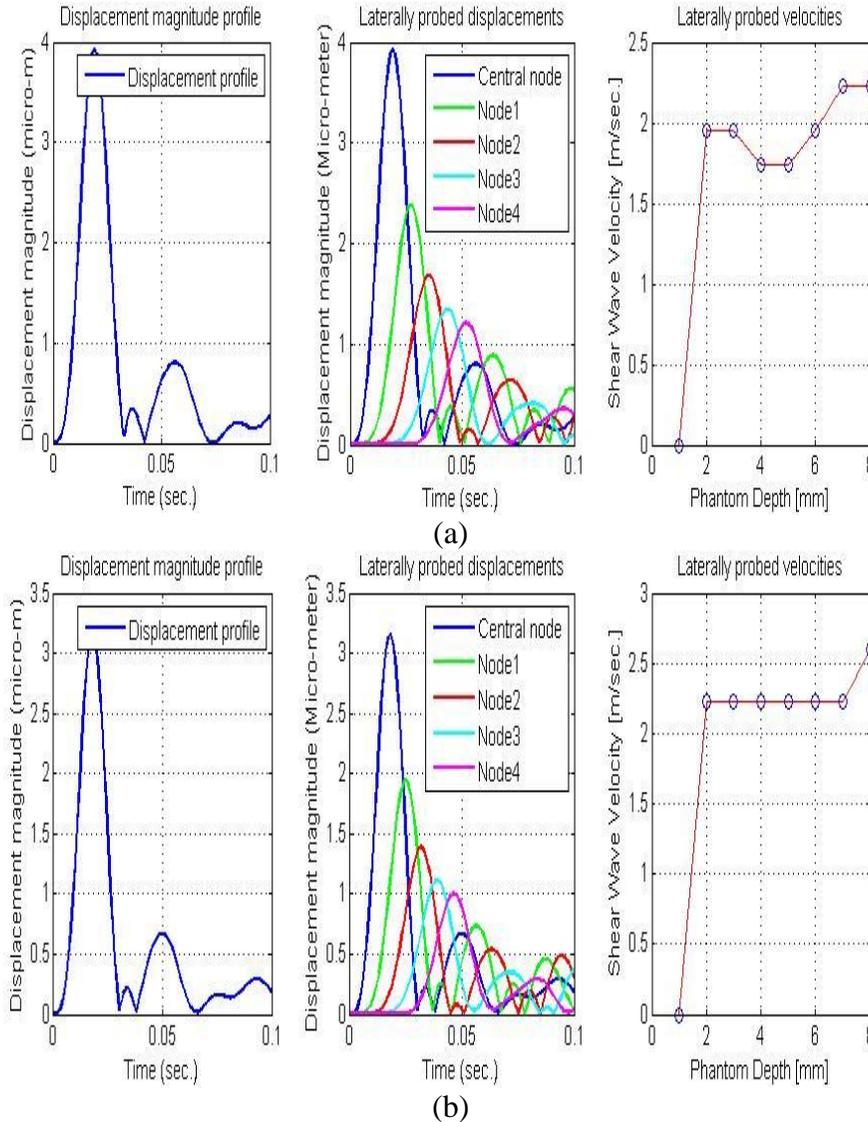

(a)

(b)

Figure 5.4: Resulting Shear Wave Speeds for normal fatty breast model where; the left hand side curve illustrate the displacement magnitude profile, the middle curve illustrate the displacement magnitudes at five probed nodes and the right hand side curve illustrate the velocity curve versus the laterally probed nodes. (a) Lower elastic modulus value. (b) Upper elastic modulus value.

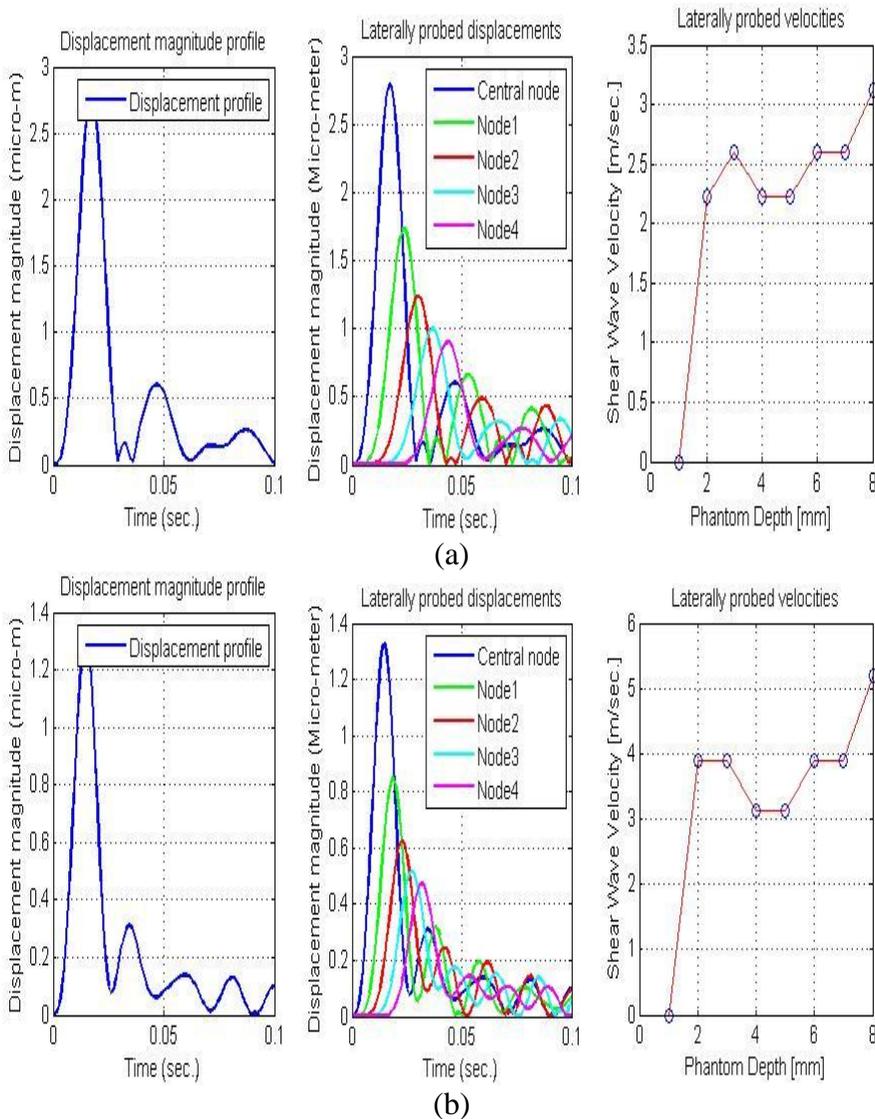

Figure 5.5: Resulting Shear Wave Speeds for normal glandular breast model where; the left hand side curve illustrate the displacement magnitude profile, the middle curve illustrate the displacement magnitudes at five probed nodes and the right hand side curve illustrate the velocity curve versus the laterally probed nodes. (a) Lower elastic modulus value. (b) Upper elastic modulus value.

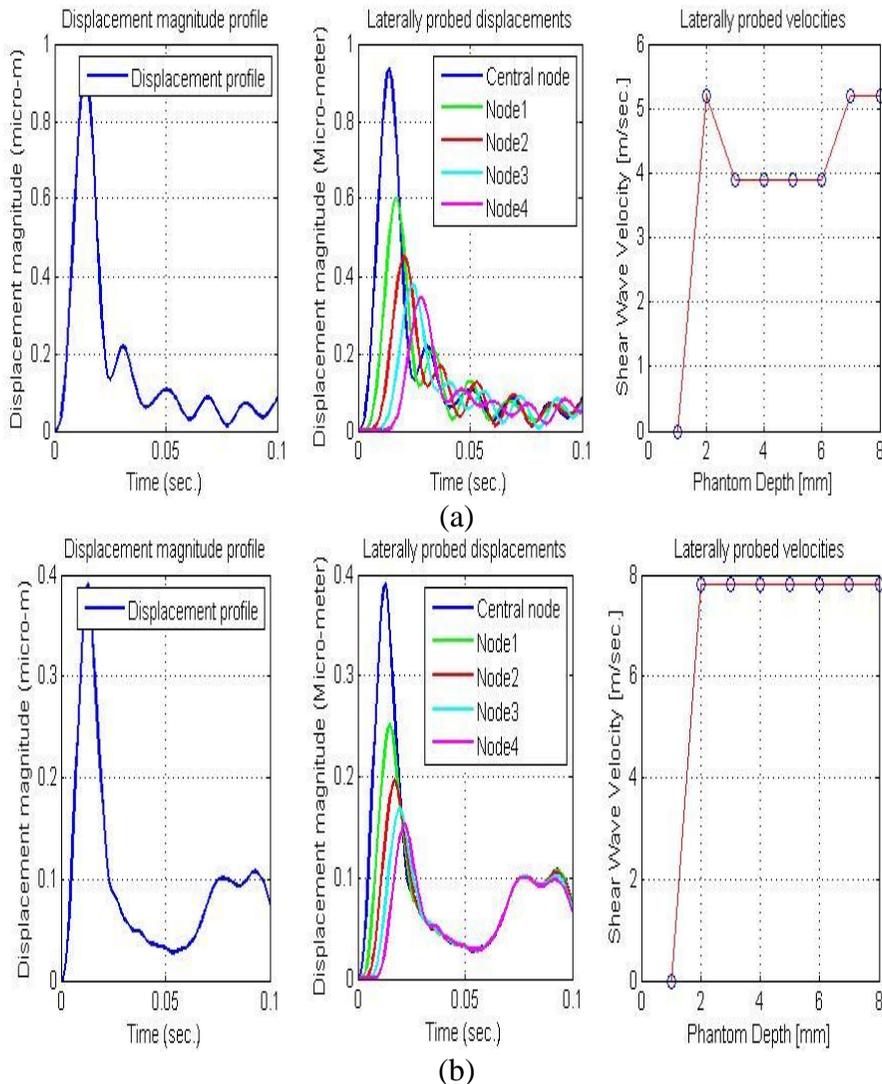

(a)

(b)

Figure 5.6: Resulting Shear Wave Speeds for fibrous glandular breast model where; the left hand side curve illustrate the displacement magnitude profile, the middle curve illustrate the displacement magnitudes at five probed nodes and the right hand side curve illustrate the velocity curve versus the laterally probed nodes. (a) Lower elastic modulus value. (b) Upper elastic modulus value.

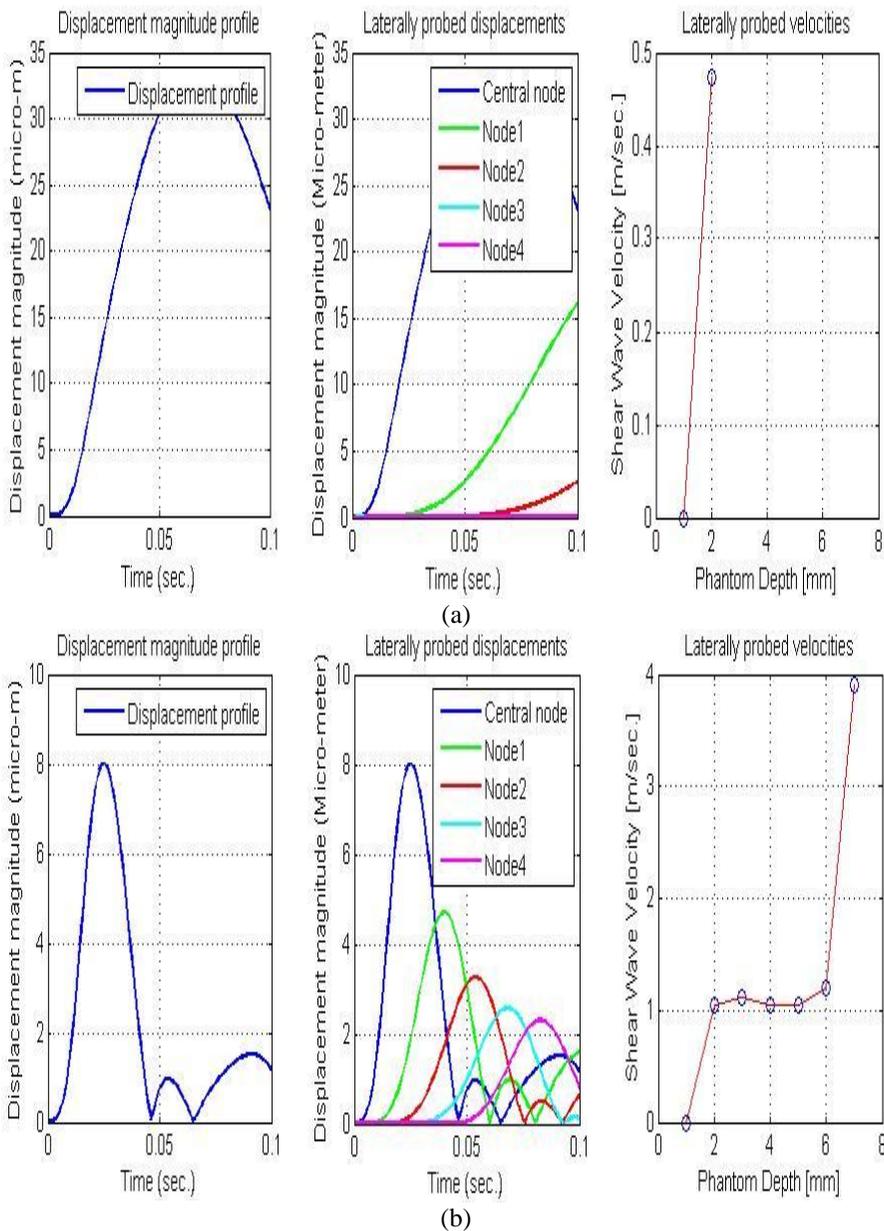

Figure 5.7: Resulting Shear Wave Speeds for normal liver model where; the left hand side curve illustrate the displacement magnitude profile, the middle curve illustrate the displacement magnitudes at five probed nodes and the right hand side curve illustrate the velocity curve versus the laterally probed nodes. (a) Lower elastic modulus value. (b) Upper elastic modulus value.

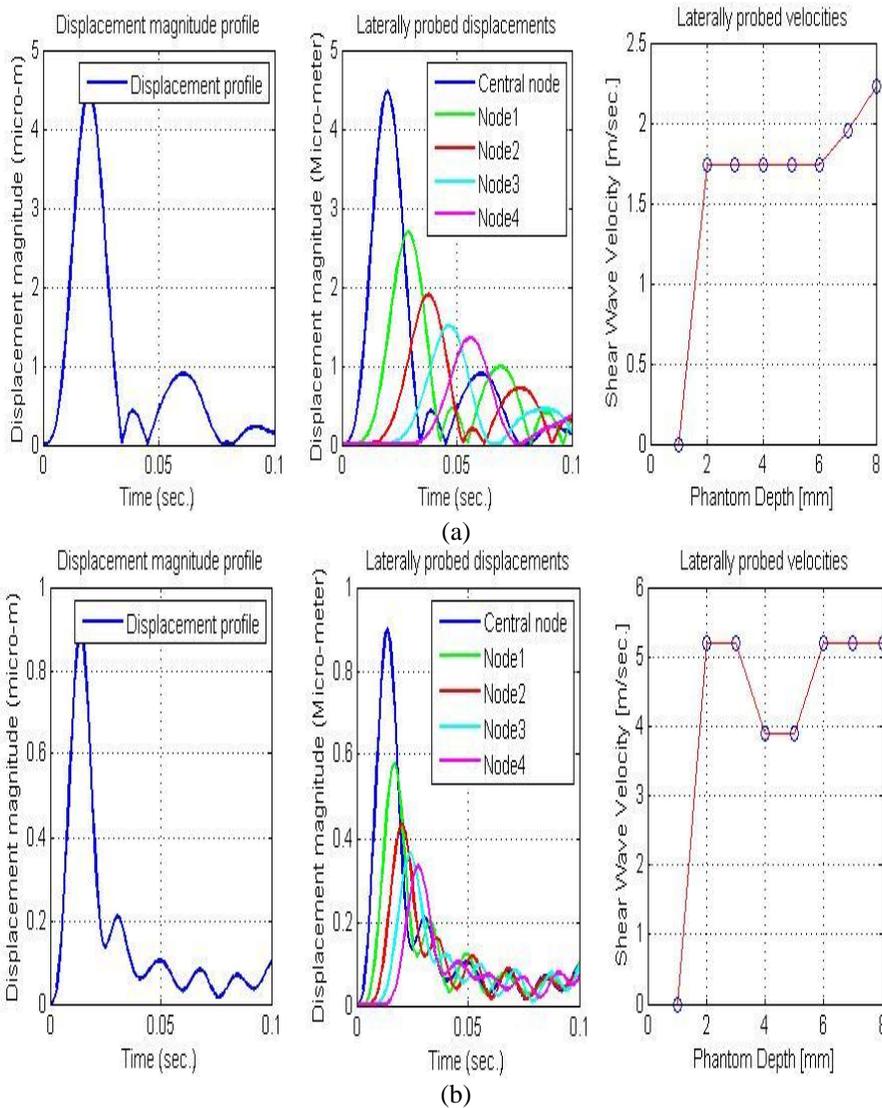

(a)

(b)

Figure 5.8: Resulting Shear Wave Speeds for cirrhosis liver model where; the left hand side curve illustrate the displacement magnitude profile, the middle curve illustrate the displacement magnitudes at five probed nodes and the right hand side curve illustrate the velocity curve versus the laterally probed nodes. (a) Lower elastic modulus value. (b) Upper elastic modulus value.

Table 5.4: Estimated and calculated shear wave speed values for phantom models used in case studies of breast and liver tissues.

| Phantom | Average Shear Wave Speed (m/sec.) (Estimated) | Average Shear Wave Speed (m/sec.) (Calculated) | Accuracy % of estimated speed w.r.t the calculated speed |
|---|---|---|---|
| Normal fatty breast tissue | 2.0018-2.2853 | 2.3791-2.7472 | 83.2%-84.3% |
| Normal glandular breast tissue | 2.5191-3.9807 | 2.9673-4.5557 | 84.7%-87.4% |
| Fibrous glandular breast tissue | 4.8363-7.4405 | 5.4944-8.7595 | 87.9%-85.0% |
| Normal liver tissue | 0.4735-1.5179 | 0.3546-1.3736 | ------------ |
| Cirrhosis liver tissue | 1.8380-5.0223 | 2.1718-5.6077 | 84.3%-89.6% |

### 5.3.3. Discussion

The shift in shear wave estimation for both the breast and the liver models arises due to the lack in the number of nodes involved in the construction of the FEM Mesh. Yet the estimation is done at an optimum frame rate of about 1 KHz.

Displacement magnitude profiles hold valuable information about the tissue's elastic properties. This is because of the ability to investigate the Time-To-Peak (TTP) displacement profile of any node clearly. Yet, the central node (the node at which the focal point of the ARFI is located) is much valuable than any other node in homogeneous FEMs, as it gives more data about the model behaviour with ARFI versus time of investigation.

The displacement magnitude profile in Fig. 5.2 (a and b) shows clearly that the normal fatty breast tissue tends to have a very low shear wave speed value. This is evident at the second highest peak in both curves. At the lower elastic moduli value model, this second highest peak tends to reach its peak displacement after 0.05 seconds, the upper elastic moduli value model tends to have the its peak displacement at 0.05 seconds typically.

In normal glandular; as in Fig. 5.3 (a & b) this peak is shifted to the left, reaching its peak displacement before 0.05 seconds. In other words, 0.05 seconds is a time mark and this peak can be considered as a yardstick to differentiate between normal fatty breast and normal glandular breast. It is a landmark before which the breast tissue is considered as normal glandular tissue and after which the breast tissue is considered as normal fatty tissue. Increase in the stiffness of the breast tissue yields faster shear wave, i.e. higher shear wave speed. Also, higher stiffness yields damped shear wave peaks, as seen in Fig. 5.4.b, where the second, third and fourth wave peaks that are evident in Fig. 5.4.a are damped and superimposed to form the last peak shown at the distal end of the right hand side in Fig. 5.4.

The displacement magnitude profile in Fig. 5.5.a shows a delayed response for the normal liver tissue model leading to a very slow shear wave speed. As the elastic modulus of this tissue model increases the tissue response becomes faster, this is shown in Fig. 5.5.b and Fig. 5.6.b. Also, more peaks are formed, the second highest one; again; is the most important one, as it differentiates between normal and abnormal tissue behaviours.

The second highest peak location with respect to time, typically the 0.05 seconds landmark, can be used to specifiy the tissues' relative elasticity. Low elastic moduli tissues tend to have this peak far at the right hand side of the curve, while higher elastic moduli tissues tend to have it at the left hand side of the curve. As this peak moves to the left of the 0.05 secods landmark, the tissue becomes more and more stiffer, the shear wave speed becomes faster as well.

In liver tissue model, another instance is noticed, the second peak, the intermediate peak between the first highest peak and the second highest peak, can be used to differentiate between normal liver tissue and liver cirrhosis. This is seen in Fig. 5.5.b and Fig. 5.6.a, where Fig. 5.5.b specifies the upper margin of the normal liver, and at this margin the second peak occurs at approximatly 0.06 seconds, whereas at the lower margin of liver cirrhosis illustrated in Fig. 5.6.a, this peak occurs at approximatly 0.035 seconds. A time margin of 0.025 seconds separates the normal and cirrhosis liver cases.

## 5.4.  Summary

The 2nd highest peak in decreasing order appearing in the displacement magnitude curve of the central node is related to the stiffness of the tissue under investigation.

For tissues of low stiffness this peak is occurring after the time mark 0.05 seconds. And for higher stiffness tissues this peak tends to occur earlier before 0.05 seconds time mark. The time at which this peak occurs along with the estimated shear wave speed can be used as a metric for describing the tissues' elastic status.

In this chapter, five experiments with five different elastic moduli phantoms are performed so that the relationship of the $2^{nd}$ highest peak; appearing in the displacement magnitude profile; with the stiffness is investigated. Also this conclusion is further used as a metric for describing the stiffness of the case studies; i.e. for breast and liver tissues. Experiments for case studies are performed to investigate the availability of using this metric to differentiate between the different cases of the tissues.

# Chapter 6 | Conclusion and Future Work

## 6.1. Conclusion

In this thesis a FEM simulation is used to study the behavior of the soft tissue to a locally induced displacement by the acoustic radiation force impulse. Two dimensional finite element strain model is constructed to simulate the soft tissue mechanical properties, also to simulate the behavior of soft tissue to acoustic radiation force impulse. Acoustic force is applied for a very short duration, i.e. transiently, to induce local deformation to the model nodes.

Hereby, generating a propagating wave formed from locally induced displacements. The speed of the generated wave is related to stiffness of the corresponding model. Generalized model for soft tissue; Agar-gelatine model; is used to simulate the soft tissue due to the ability of the Agar to simulate scatterers in tissues and the Gelatine to hold the elasticity and viscosity of the model. Acoustic radiation force impulse is generated and applied to the focal point of the ultrasound beam to induce the local displacement inside the model. The resulting shear wave is then estimated from eight fixed lateral positions using TTP method. For Agar-gelatin phantom the shear wave speed (SWS) was estimated to be 5.2 m/sec. compared to 5.7 m/sec. which is a good estimation for the speed.

The relationship of displacement magnitude profile with the stiffness value is investigated as well by means of preliminary experiments. Five phantoms are constructed and their mechanical properties are set such that they resemble various soft tissue cases. As a concluding remark from these experiments the shear wave speeds are estimated for these five phantoms; typically 1.13, 1.46, 2.28, 3.16 and 3.86 m/sec. A very interesting conclusion from these experiments is that the $2^{nd}$ highest peak present in the displacement magnitude profile occurs after the time mark 0.05 seconds for lower stiffness phantoms and before this time mark for higher stiffness phantoms, i.e. the lower speeds and higher speeds respectively. This remark along with the estimated shear wave speed can be used to distinguish the status of the tissue; whether it is healthy or pathologic one.

Two case studies for the breast and liver tissues have been investigated. Three tissue states for the breast tissue have been studied, namely the normal fatty, the normal glandular and the fibrotic glandular breast tissues. Their shear wave speed values have been estimated as well, and found to be (as a range of lower and upper edge for each state) as follows: 2.00-2.28 m/sec, 2.51-3.98 m/sec., and 4.83-7.44 m/sec. for normal fatty tissue, normal glandular tissue and fibrotic glandular breast tissue respectively. Two tissue states of liver tissue have been investigated as well, the normal liver tissue and cirrhosis liver tissue. Also, their shear wave speed ranges were found to be 0.47-1.51 m/sec. and 1.83-5.02 m/sec. for normal and cirrhosis respectively.

Parameters related to ARFI and affected by the shift in stiffness are investigated as well, where amongst these parameters the peak displacement and time-to-peak displacement. Time-To-Peak displacement is found to be decreasing by the increase in the stiffness value at the same ARFI and same excitation frequency. This is evident from the curves due to the dashpot element effect. The peak displacement is found to be decreasing by the increase of the stiffness value as well. And again due to the dashpot element effect which responds slowly and poorly to rapid changes and higher stiffness values. Resolution of the tracking transducer is found to be related to the stiffness of the investigated tissue, where higher stiffness tissues require higher resolution transducers to track the generated wave accurately. Frame rate used in tracking the generated wave is related also to the stiffness of the tissue, as the increase in stiffness requires higher frame rate to track the actual wave induced inside the tissue.

## 6.2. Future work

Finishing this thesis with highlighting the points that can be added for further work may improve the search in this field at investigating them in more details. These points are listed in brief as follows:

➢ Increasing the number of involved nodes constituting the two dimensional strain model, as the increase in their number leads to an accurate estimation of the tracked shear wave  because these nodes forms the stationary locations of tracking.

➢ Investigating the effect of fitting the resulting displacement magnitude curve to have the Gaussian and/or Hilbert shape for leading to a more precise estimation of the wave speed.

➢  Studying the effect of the stiffness value shift on the parameters related to the ARFI mentioned in chapter 6 using a real ultrasound transducer away from the simulation to conclude how close the experimental results are consistent with the clinical results.

➢ Applying the work concluded in this research on real phantoms and clinical experiments.

➢ Applying the concluded work in this thesis and from the preceding points on a glaucomatous eye model to develop a method to estimate the stiffness of the glaucomatous eye precisely using the ocular ultrasound imaging techniques.

➢ Finally, investigating the non-homogenous tissues, where the inclusions are present, to study the behavior of these tissues to a transient force.

# Appendix

## Appendix A

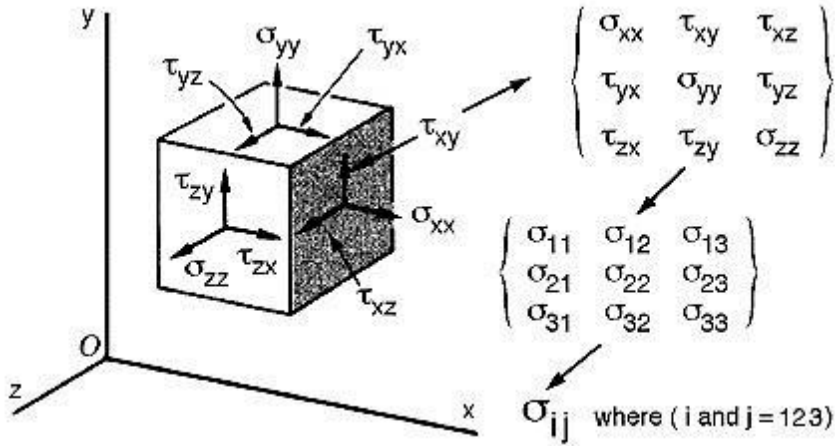

Figure A.1: Stress tensor [5].

Now, $\mathbf{C_{ijkl}}$ is further discussed and its proof is explained in details for the elastic material. $\mathbf{C_{ijkl}}$ is called the $4^{th}$ order stiffness tensor, and its stiffness matrix will be reduced to 6 x 6 stiffness matrix where this matrix relates six components of symmetric stress tensor $\boldsymbol{\sigma_\alpha}$ to a six components of a symmetric strain tensor $\boldsymbol{\epsilon_\beta}$ as shown in Fig. A.1.

The general form of $\mathbf{C_{ijkl}}$ states $\boldsymbol{\sigma_{ij}}$ as stress and $\boldsymbol{\epsilon_{kl}}$ as **lagrangian strain**, $\boldsymbol{\sigma_{ij}} = \mathbf{C_{ijkl}}\boldsymbol{\epsilon_{kl}} = \mathbf{C_{ijkl}l_{kl}}$ . For the feasibility we investigate how the stress is written as a $2^{nd}$ order tensor, and by symmetry we write the strain.

Here, $\boldsymbol{\sigma_{ij}}$ is the normal stress component and $\boldsymbol{\tau_{ij}}$ is the shear stress component.

But here arises a problem, where writing the $2^{nd}$ order strain tensor is not as straight forward as it was for the stress tensor, as each stress component could be seen simply as a force acting on the surface of a cube.

Strain implies displacement which requires much more development of the equation, where for non-linear strain displacement the relationship is given by:

$$L_{ij} = \frac{1}{2}\left(u_{ij} + u_{ji} + u_{ri}u_{ir}\right) \qquad (A.1)$$

Where the $u_{ij}$ represents the potential function.

Assuming a very small displacement gradients, so the last term in the equation is neglected as it represents the non-linear large displacement gradients. Also, the upper case **L** is reduced to the lower case **l** denoting the **lagrangian** strain tensor, Eqn. A.1 becomes:

$$l_{ij} = \frac{1}{2}\left(u_{ij} + u_{ji}\right) \qquad (A.2)$$

Now; with the stress and strain defined as a $2^{nd}$ order tensors, it is not difficult to derive the desired linear isotropic relationship ($\boldsymbol{\sigma_{ij} = C_{ijkl}\epsilon_{kl}}$). From the $1^{st}$ law of thermodynamics where one of its forms is given by:

$$\frac{Du}{dt} = \frac{1}{\rho}\left(\sigma_{ij}\right)\frac{d}{dt}\left(\epsilon_{kl}\right) - \frac{1}{\rho}\left(b_{ij}\right) + C \qquad (A.3)$$

Where:
1. **u** is the specific internal energy per mass unit.
2. **ρ** is the density of the material.
3. $\boldsymbol{\epsilon_{kl}}$ is the Eulerian strain associated with the Eulerian displacement gradient.
4. **C** is the radiative heat transfer per unit time per unit mass.
5. $\mathbf{b_i}$ is the heat transferred through a unit surface per unit time.

If we assume no heat transfer by radiation or conduction, and we assume that only small displacement gradients result in **lagrangian strains $l_{ij}$**. Eqn. 15 will be reduced to a form where **u** is now called the **strain energy per unit mass** since the only mechanism available for storing internal energy in the continuum is through deformation. Hence, Eqn. A.3 is reduced to:

$$\frac{Du}{dt} = \frac{1}{\rho} \left( \sigma_{ij} \right) \frac{d}{dt} \left( \epsilon_{kl} \right) \tag{A.4}$$

Next we assume $\boldsymbol{\rho}$ is constant or at least a function of strains, and also we assume stresses are related to strains by some unknown function to be determined later. Hence for the purpose of our discussion; the strain energy can be written as a function of strain and given by:

$$U = u(u_{11} + u_{22} + \cdots + u_{33}) \tag{A.5}$$

From Eqn. A.5 and by parial differentiation, this leads to:

$$Du = \frac{\partial u}{\partial l_{ij}} \, dl_{ij} \tag{A.6}$$

By substituting by (A.6) in (A.4) we obtain Eqn. A.7.

$$\left( \frac{\partial u}{\partial l_{ij}} - \frac{\sigma_{ij}}{\rho} \right) dl_{ij} = zero \tag{A.7}$$

And for any arbitrary non-zero $\boldsymbol{dl_{ij}}$, the following equation, Eqn. A.8 is obtained, and this is the relation between stress, strain and strain energy per unit mass.

$$\sigma_{ij} = \rho \, \frac{\partial u}{\partial l_{ij}} \tag{A.8}$$

But, we assumed before in the preceding literature at some point that there is an undefined functional form for strain energy as a function of strain. There is one possible function form would be to expand the strain energy into a power series whose coefficients; like strain; are tensors but constants with indices that sum with the indices of the strain tensors such that each term in the series is a scalar component of the strain energy. This is given by:

$$U = \alpha + \beta_{ij}l_{ij} + C_{ijkl}l_{ij}l_{kl} + \delta_{ijklmn}l_{ij}l_{kl}l_{mn} + \cdots \tag{A.9}$$

The 1$^{st}$ term is an arbitrary energy reference state (we can set it to zero).

The 2$^{nd}$ term is a zero strain reference state.

The 3$^{rd}$ term is a linear elastic term.

The 4$^{th}$ term is a non-linear elastic term.

We can add higher order terms as needed to model the desired constitutive stress-strain relationship.

By substituting by Eqn. A.8 in Eqn. A.9 we obtain the most general form of the isotropic elastic constitutive law in tensor format, and given by:

$$\sigma_{ij} = \rho\beta_{ij} + \rho\, C_{ijkl}l_{kl} + \delta_{ijklmn}l_{kl}l_{mn} + \cdots \qquad (A.10)$$

Where:

1. $\boldsymbol{\beta_{ij}}$ represents the residual stresses when $\mathbf{l_{ij}}$ goes to zero.

2. $\boldsymbol{\delta_{ijklmn}}$ represents non-linear elastic term giving rise to non-linear stress-strain diagrams, and always set to zero for linear elastic models.

Now the equation becomes as following:

$$\sigma_{ij} = \rho C_{ijkl}l_{kl} \qquad (A.11)$$

And for strain per unit volume, the equation is reduced to:

$$\sigma_{ij} = C_{ijkl}l_{kl} \qquad (A.12)$$

Hence, the proof of the generalized Hook's law for tensor form of the elastic material has come to an end.

# Appendix B

In the following part further decomposition of both the stress and strain tensors, and the contribution of each constituent will be investigated. Each of the stress and the strain can be decomposed into **deviatoric** and **dilatational** parts. For the stress tensor [4], the equation will be:

$$\sigma_{ij} = \sigma'_{ij} + \frac{1}{3}\left(\sigma_{kk}\delta_{ij}\right) = \sigma'_{ij} + \sigma_H\delta_{ij} \qquad (B.1)$$

Similarly for the strain, the equation will be:

$$\epsilon_{ij} = \epsilon'_{ij} + \frac{1}{3}\left(\epsilon_{kk}\delta_{ij}\right) = \epsilon'_{ij} + \frac{1}{3}\left(\epsilon_v\delta_{ij}\right) \qquad (B.2)$$

Moreover, we can define the relationships between both the deviatoric stresses and strains, and the dilatational stresses and strains, to be given by Eqn. 21 and 22.

$$\sigma_{ij} = 2G\,\epsilon'_{ij} \qquad (B.3)$$

$$\sigma_H = K\,\epsilon_v \qquad (B.4)$$

Where **K** and **G** are given by:

$$\mu = \frac{E}{2\,(1+v)} = G \qquad (B.5)$$

$$K = \frac{E}{3(1-2v)} \qquad (B.6)$$

# Appendix C

This section discusses the effect of shifting the stiffness of soft tissue on some parameters obtained during the elastography imaging procedure. Amongst these parameters, the Time-To-Peak displacement and the axial peak displacement at the central node are introduced.

**Time-To-Peak displacement versus stiffness of soft tissue**

Time-To-Peak displacement curves for central nodes for phantoms introduced in preliminary experiments and introduced in case studies are obtained and the time at which the peak occurs for each curve is plotted versus its stiffness. Having TTP values plotted versus a variety of corresponding stiffness values; the TTP versus stiffness curve is completely constructed. The curve is shown in Fig. C.1.

The TTP decreases as the stiffness of the soft tissue increase and can be modeled as a decaying exponential function as shown in Fig. C.1.
This decaying effect in the excitation time or the time taken to reach the peak displacement for the central node arises due to the effect of the dashpot element present in Maxwell model. This dashpot element responds well to the low frequencies of excitation, and responds very slowly to high frequency excitations, in case of the same stiffness value. Also, dashpot responds slower to the same frequency at shifting the stiffness of the model towards higher stiffness values.

The decrease in the TTP value has a role in increasing the frame of imaging the migrating shear wave speed, where precise tracking of this wave comprises tracking its peak accurately. High frame rate ensures that the tracking procedure is actually tracking the real peak of the propagating wave not the peak captured by the tracking beam.

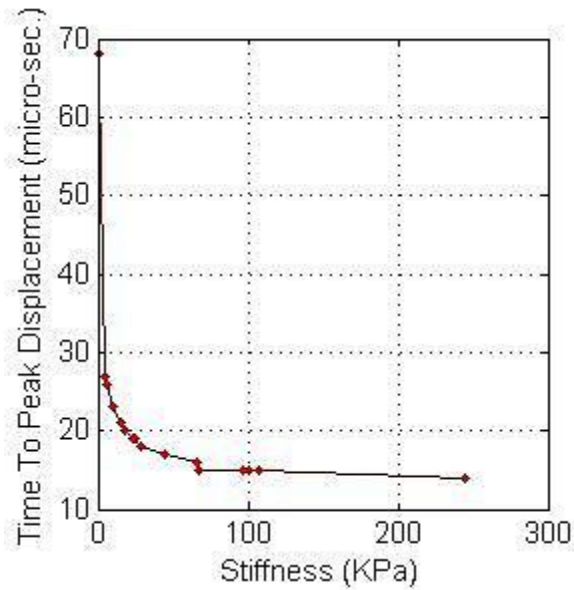

Figure C.1: TTP versus stiffness of soft tissue.

**Peak Displacement versus Stiffness of soft tissue**

Peak displacement values for phantoms used in preliminary experiments and case studies are acquired. Peak displacement is the magnitude of the displacement of a specific node in the three directions x-, y- and z-direction.

The peak displacement value of the central node for each stiffness value is obtained and plotted versus its stiffness. Clearly, the peak displacement of the central node; and any node; is having a decreasing behavior with the increase of the stiffness at the same excitation frequency, as shown in Fig. C.2.

This decreasing behavior is due to the dashpot element effect, just as mentioned earlier. The decrease in the peak displacement occurs because the node under investigation cannot keep up with the frequency of excitation; which is not altered or changed; at higher stiffness values. Thus, smaller displacement is achieved at higher stiffness values. Larger displacement values are observed at smaller stiffness values, where nodes under investigation can reach its peak displacement (deformation) easily after excitation as the dashpot element responds well enough for the same excitation frequency at smaller values of stiffness.

A trade-off arises between the peak displacement and the resolution of the transducer used in tracking the propagating wave. As the peak displacement of the model nodes' becomes smaller and smaller, their tracking process becomes more difficult. This is due to the resolution of the tracking transducer is fixed at some axial and lateral resolution values but the peak displacements tracked are variable, and their variation is related to the stiffness of the soft tissue, as shown in Fig. C.2. Hence, there is an optimum resolution; axially and laterally; for tracking the propagating shear wave at each stiffness value. The limitation here is in an increasing order, which means that the transducer used to track the resulting wave for 5.2 KPa is not sufficient enough to track the resulting one for 9.8 KPa or for 23.9 KPa and so on. Instead it is sufficient for tracking the shear waves propagating in 5.2 KPa and lower soft tissues stiffness values, i.e. 4 KPa and 0.4 KPa.

The insufficient tracking results in an underestimation of the tracked wave, leading to estimating a value of the wave less than its real value.

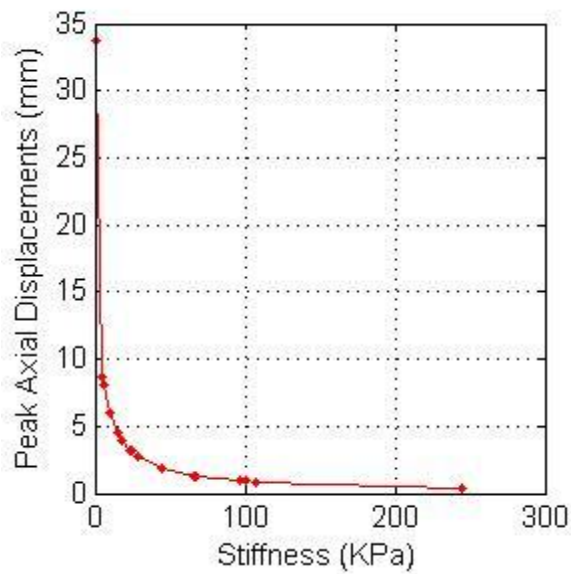

Figure C.2: Peak displacements versus stiffness of soft tissue.